\documentclass[preprint,authoryear,3p]{elsarticle}

\usepackage{natbib}

\usepackage{bm}
\usepackage{cite}
\usepackage{amsmath,amssymb,amsfonts}
\usepackage{algorithm,algpseudocode}
\usepackage{graphicx}
\usepackage{hyperref}
\usepackage{textcomp}
\usepackage{multirow}
\usepackage{xcolor}
\usepackage{colortbl}
\usepackage[export]{adjustbox}

\journal{Medical Image Analysis}

\begin{document}
\begin{frontmatter}



\title{A Plug-and-Play Method for Guided Multi-contrast MRI Reconstruction based on Content/Style Modeling} 


\author[lumc]{Chinmay Rao\corref{cor1}}
\author[lumc]{Matthias van Osch}
\author[cardiologs,tue]{Nicola Pezzotti}
\author[lumc]{Jeroen de Bresser}
\author[lumc]{Mark van Buchem}
\author[lumc]{Laurens Beljaards}
\author[hamburg]{Jakob Meineke}
\author[best]{Elwin de Weerdt}
\author[lumc]{Huangling Lu}
\author[hamburg]{Mariya Doneva}
\author[lumc]{Marius Staring}

\affiliation[lumc]{organization={Department of Radiology, Leiden University Medical Center},
 city={Leiden}, country={The Netherlands}}
\affiliation[cardiologs]{organization={Cardiologs}, city={Paris}, country={France}}
\affiliation[tue]{organization={ Department of Mathematics and Computer Science, Eindhoven University of Technology}, city={Eindhoven}, country={The Netherlands}}
\affiliation[hamburg]{organization={Philips Innovative Technologies}, city={Hamburg}, country={Germany}}
\affiliation[best]{organization={Philips}, city={Best}, country={The Netherlands}}

\cortext[cor1]{c.s.rao@lumc.nl}

\begin{abstract}
Since the various contrast-weighted MR images of a given anatomy contain redundant information, one contrast can be used to guide the reconstruction of another undersampled contrast acquired subsequently in the same session. To solve this reconstruction problem leveraging multi-contrast side information, several end-to-end learning-based guided reconstruction methods have been proposed. However, a key challenge is the requirement for large paired training datasets comprising raw k-space data and aligned reference images.
We propose a modular plug-and-play approach, which requires no k-space training data and relies solely on partially paired image-domain datasets.
In this approach, a content/style model of two-contrast MR data is first learned from a purely image-domain dataset and subsequently applied as a plug-and-play operator in iterative reconstruction. The disentanglement of content and style allows explicit representation of contrast-independent and contrast-specific factors. Consequently, incorporating prior information into the reconstruction reduces to a simple replacement operation on the aliased content of the estimated image using high-quality content derived from the reference scan. Combining this so-called \textit{content consistency} operation with an MR data consistency step, followed by a corrective procedure for the content estimate, yields an iterative scheme. 
We name this novel approach \textit{PnP-CoSMo}. 
This approach, by design, offers cross-contrast generalizability and provides an explanatory framework based on the shared and non-shared generative factors underlying the two given contrasts. We explore various aspects of PnP-CoSMo, including interpretability and convergence, via simulations. Furthermore, its practicality is demonstrated on the public NYU fastMRI DICOM dataset, showing equivalent or superior quality and greater generalizability compared to end-to-end methods.
On two in-house multi-coil datasets, PnP-CoSMo enabled up to 32.6\% greater acceleration over non-guided plug-and-play reconstruction at given SSIM.
\end{abstract}




\end{frontmatter}

\section{Introduction}
\label{sec:introduction}
\noindent
Magnetic resonance imaging (MRI) is an invaluable medical imaging modality due to the rich anatomical information it provides, with the absence of radiation-related risks, e.g., in contrast to computed tomography (CT), leading to its wide application in clinical practice. However, its central limitation is the inherently slow data acquisition process. The raw sensor data is acquired in the frequency domain (k-space), from which the image is reconstructed. Over the last 25 years, advancements such as parallel imaging \citep{pruessmann2001advances,griswold2002generalized}, compressed sensing (CS) \citep{lustig2007sparse}, and deep learning reconstruction \citep{aggarwal2018modl,pezzotti_adaptive_2020} have enabled considerable speedups by allowing sub-Nyquist k-space sampling complemented by computationally sophisticated reconstruction. These techniques have subsequently been implemented on commercial MRI systems and have demonstrated (potential) improvements in clinical workflow \citep{seow2024evaluation}.

A clinical MRI session typically involves acquiring multiple contrast-weighted scans of the same anatomy through the application of different MR pulse sequences. Since these scans are different reflections of the same underlying reality, they share a high degree of shared structure. However, currently deployed clinical protocols acquire and reconstruct each scan as an independent measurement, ignoring the information redundancy across scans. There is, therefore, an opportunity to further optimize acquisition times of MRI sessions by exploiting this shared information. On the reconstruction side, multi-contrast methods have addressed this problem by introducing shared information into the reconstruction phase, enabling higher levels of k-space undersampling. In the simplest case with two contrasts, multi-contrast reconstruction can be classified into two types: (a) guided reconstruction, where an existing high-quality reference scan is used to guide the reconstruction of an undersampled second scan \citep{ehrhardt_multicontrast_2016,weizman2016reference,zhou_dudornet_2020} and (b) joint reconstruction, where both contrasts are undersampled and are reconstructed simultaneously \citep{bilgic2011multi,huang2014fast,kopanoglu2020simultaneous}. In this work, we consider the problem of guided reconstruction, assuming no inter-scan motion between reference and target scans.

\begin{figure*}
    \centering
    \includegraphics[width=1\textwidth]{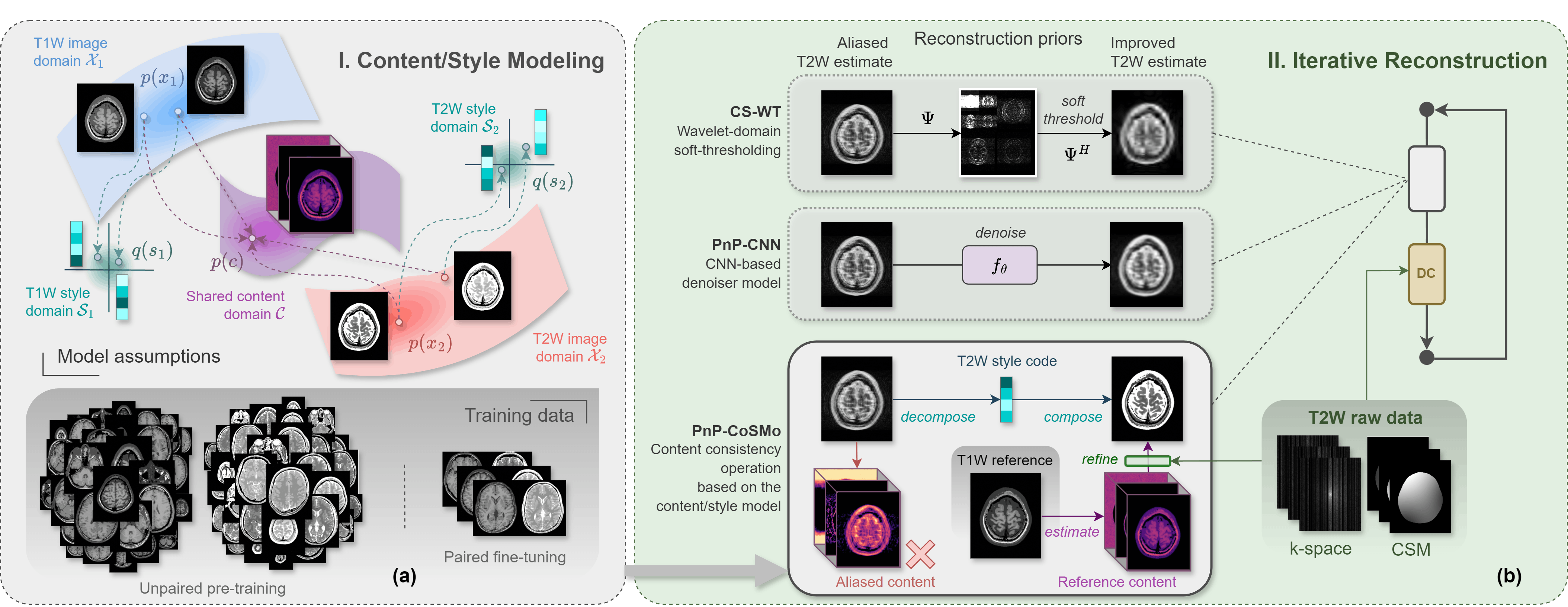}
    \caption{Our two-stage approach to guided reconstruction. (a) The first stage learns a content/style model of two-contrast MR image data offline. No k-space data is required in this stage. We assume that the two image domains $\mathcal{X}_1$ and $\mathcal{X}_2$ can be decomposed into a shared content domain $\mathcal{C}$ and separate style domains $\mathcal{S}_1$ and $\mathcal{S}_2$. This model is learned in two stages: an unpaired pre-training stage and a paired fine-tuning stage, both requiring only image data. (b) The reconstruction stage applies the content/style model as a \textit{content consistency operator} (bottom) within an ISTA-based iterative scheme. Given an aligned reference image, guidance is introduced into the reconstruction by simply replacing its aliased content with content derived from the reference. The "refine" block denotes a content refinement update, which iteratively corrects for inconsistencies between the reference content and the measured k-space data, improving the effectiveness of the content consistency operator. DC denotes data consistency. For comparison, two other reconstruction priors are shown, namely wavelet-domain soft-thresholding (top) and CNN-based denoising (middle) used in CS-WT and PnP-CNN algorithms, respectively.}
    \label{fig:overview}
\end{figure*}

The guided reconstruction problem entails using the local structure of the reference scan as a prior to complement the undersampled k-space measurements of the target scan. This problem has been formulated in different ways, ranging from conventional CS reconstruction \citep{ehrhardt_multicontrast_2016,weizman2016reference} to end-to-end learning with unrolled networks \citep{zhou_dudornet_2020,yang2020model,pooja2022mc,lei2023decomposition} and, more recently, diffusion model-based Bayesian maximum \textit{a posteriori} estimation \citep{levac2023mri}. Most end-to-end approaches, although more powerful than earlier hand-crafted ones, suffer from the main drawback of requiring large paired training datasets consisting of the target image and its k-space together with an aligned reference image, thereby limiting their application on real-world MR data. We address this issue by proposing a plug-and-play reconstruction method that splits the problem into a purely image-domain learning sub-problem and an iterative reconstruction sub-problem. The learning problem leverages ideas from disentangled content/style representation learning, thereby offering greater interpretability to our approach. To the best of our knowledge, the present work is the first to thoroughly explore content/style decomposition for multi-contrast MRI reconstruction.

In recent years, image-to-image translation has been applied in the direct estimation of one MR contrast from another \citep{dar_image_2019,yurt_mustgan_2021,denck_mr-contrast-aware_2021,atli2024i2i}. While these methods are attractive due to their lack of dependence on k-space data, cross-contrast translation takes an extreme stance, when viewed in the light of MR physics, by not explicitly taking into account contrast-specific sensitivities and relying solely on the prior contrast. Hence, in the context of MR image formation, mere cross-contrast translation can only provide a part of the information about the target image, since data acquisition via MR sequences is necessary to obtain new information about the anatomy. That being said, literature on unpaired image translation provides a repertoire of useful tools such as joint generative modeling of two-domain image data \citep{liu2017unsupervised,huang2018multimodal}, which can be adopted to complement image reconstruction. We observe that methods such as MUNIT \citep{huang2018multimodal} can be applied to learn semantically meaningful representations of contrast-independent and contrast-specific information as content and style, respectively, without the need for paired image-domain training data.

Plug-and-play (PnP) methods are an emerging paradigm for solving inverse problems in computational imaging. The main research line \citep{ahmad_plug-and-play_2020,kamilov_plug-and-play_2023} has focused on learning convolutional neural network (CNN) based denoising models on image-domain data and applying them as functions that replace proximal operators in iterative algorithms, such as the iterative soft-thresholding algorithm (ISTA) and the alternating direction method of multipliers (ADMM). In addition to an improved image recovery, an advantage of this approach is the decoupling of the learning problem of image modeling from the inverse problem of image reconstruction, thereby simplifying model training and improving generalizability across different acceleration factors, undersampling patterns, etc. With this design pattern in mind, we combine content/style-based image modeling and iterative reconstruction under a PnP-like framework.

We first leverage semantic content/style modeling to learn explicit representations of contrast-independent and contrast-specific components from two-contrast MR image data. This training process is independent of the reconstruction problem and can be performed using partially-paired and even unpaired images. We then make the insightful observation that in multi-contrast MR images, the style information tends to localize in the center of k-space, enabling an accurate style estimation from an undersampled image. Since the content of the target image, which is the residual and contrast-independent information, is supplied by the reference image, the composition of a de-aliased estimate of the target image from the undersampled estimate can be performed in a single step. We term this the \textit{content consistency operation}, which forms the basis of our iterative reconstruction algorithm PnP-CoSMo (Plug-and-Play method based on Content/Style Modeling) \citep{rao2024guided}. An overview of our approach is shown in Fig. \ref{fig:overview}. While the PnP-based decoupling of image modeling and image reconstruction simplifies the training process and allows the two stages to be analyzed separately, a further level of decoupling offered by content/style disentanglement offers additional modularity and an intuitive guidance mechanism for the reconstruction. Specifically, our contributions are four-fold:
\begin{enumerate}
    \item We show that unpaired image-domain training can learn disentangled contrast-independent and contrast-specific representations, followed by a fine-tuning strategy that refines the content representation using limited amounts of paired image-domain data.
    \item With this content/style model as the basis, we define a \textit{content consistency operator} capable of removing severe undersampling artifacts from the target image, given the corresponding reference image.
    \item Extending this idea further and incorporating a corrective process for the content, we develop PnP-CoSMo, a modular algorithm for guided reconstruction combining the flexibility of the plug-and-play approach with the semantic interpretability of content/style decomposition.
    \item Through comprehensive experiments, we (a) shed light on several properties of PnP-CoSMo, such as convergence and robustness, (b) demonstrate its equivalent performance and superior generalizability compared to state-of-the-art end-to-end models, (c) demonstrate its applicability on real-world raw data, and (d) report radiological feedback based on a small-scale qualitative clinical task.
\end{enumerate}

\section{Related work}
\subsection{Reconstruction methods for accelerated MRI}
\noindent
One of the conventional paradigms for accelerating MRI by k-space undersampling is compressed sensing (CS) \citep{lustig2007sparse}, which combines random sampling with sparsity-based iterative denoising, most commonly implemented using the ISTA \citep{chambolle1998nonlinear} or ADMM (\citep{boyd2011distributed}) family of algorithms. Most modern deep learning-based reconstruction methods focus on improving the denoising component. Plug-and-play (PnP) methods \citep{ahmad_plug-and-play_2020,kamilov_plug-and-play_2023} replace the proximal operator in ISTA and ADMM with off-the-shelf denoisers such as a learned convolutional denoising model. Unrolled networks \citep{schlemper2017deep,aggarwal2018modl,pezzotti_adaptive_2020} extend this idea by casting the entire iterative algorithm into one large network, trained end-to-end. As a result, unrolled networks more adaptive to factors such as sampling pattern and acceleration, although at a cost of generalizability \citep{ahmad_plug-and-play_2020}.

One of the earliest CS-based guided reconstruction methods was based on structure-guided total variation (STV) proposed by \citet{ehrhardt_multicontrast_2016}, which assumes the sparse coefficients of the reconstruction are partially known given the high-frequency features in the reference scan. Later work introduced adaptive elements into the multi-contrast CS framework, e.g., \citet{weizman2016reference} proposed adaptive weighting-based guided CS and \citet{pingfan2018coupled} used adaptive sparse domains based on coupled dictionary learning. In the latter, the problem was formulated using a patch-level linear model comprising coupled and distinct sparse dictionary representations of the two contrasts. This model resembles a content/style model in form, although it is more restrictive. A general drawback of classical methods compared to deep learning-based ones is their lower flexibility. End-to-end learning-based methods \citep{zhou_dudornet_2020,yang2020model,liu2021optimizing,liu2021regularization,feng2022multimodal,lei2023decomposition}, on the other hand, supply the reference scan as an additional input to a deep reconstruction model, allowing it to automatically learn the suitable features to extract and transfer into the reconstruction. These methods have thus proven to be more effective than conventional algorithms. However, end-to-end methods require large paired training datasets comprising the ground-truth image and the k-space together with aligned reference images. This is an excessively strong data constraint when working with retrospectively collected clinical data that reflects the natural inconsistencies of routine practice, such as missing contrasts, differences in spatial resolutions, and the existence of inter-scan motion.
Additionally, by relying on end-to-end-learned features, end-to-end methods are generally less interpretable than their conventional counterparts in that they do not explicitly model the multi-contrast problem in terms of the underlying shared and non-shared information. MC-VarNet \citep{lei2023decomposition} is an important exception, which uses a simple linear decomposition of the reference contrast into common and unique components, applying the common component for guidance.

A reconstruction method that is as effective as the learning-based methods while having more lenient data requirements and offering a high degree of interpretability is still needed. Our plug-and-play method relies on learning a deep non-linear content/style transform from image-domain data only, even if subject-wise paired images are not fully available.

\subsection{Unpaired image-to-image modeling}
\noindent
Image-to-image modeling is the general problem of learning a mapping between two image domains and was first addressed by Pix2Pix \citep{isola2017image} and CycleGAN \citep{zhu2017unpaired} in paired and unpaired settings, respectively. Another line of unpaired image translation methods, the first of which was UNIT \citep{liu2017unsupervised}, assumes a shared latent space underlying the two domains to explicitly represent shared information. However, both CycleGAN and UNIT assume a deterministic one-to-one mapping between the two domains, ignoring the fact that an image in one domain can have multiple valid renderings in the other.
Deterministic image translation has been widely applied to MRI contrast-to-contrast synthesis \citep{dar_image_2019,oh2020unpaired,yurt_mustgan_2021}. Fundamentally, however, these methods do not account for the variability of the scanning setup that influences the realized contrast level and the differential visibility of pathologies in the target image. \citet{denck_mr-contrast-aware_2021} partially address this problem by proposing contrast-aware MR image translation, where the acquisition sequence parameters are fed into the network to control the output's contrast level. However, this model is too restrictive since it assumes a single pre-defined mode of variability (i.e., global contrast level) in the data, for which the labels (i.e., sequence parameters) must be available. 

MUNIT \citep{huang2018multimodal} extended UNIT by modeling domain-specific variability in addition to domain-independent structure, enabling many-to-many mapping and thus overcoming the rigidity of UNIT. The result was a stochastic image translation model, which, given an input image, generates a distribution of synthetic images sharing the same "content" but differing in "style". More fundamentally, MUNIT is a learned invertible transformation between the image domains and the disentangled content/style domains. And unlike other content/style modeling frameworks such as \citep{Kotovenko_2019_ICCV,Kwon_2021_ICCV,wu2023not}, MUNIT models content and style as latent generative factors of the two image domains, providing precise distribution-level definitions for them.
Clinical MR images of a given protocol contain multiple modes of variability, many of which are not known \textit{a priori}. 
Compared to \citet{denck_mr-contrast-aware_2021}, we make a broader assumption that the contrast-independent semantic information is local in nature and that the contrast-specific variations in the dataset can include global effects of acquisition settings as well as local anatomical features unique to the contrast. 
We model using an approach based on the MUNIT framework, referring to the shared and non-shared components as content and style, respectively.
 
\subsection{Combining image translation with reconstruction}
\noindent
Acknowledging the limitation of MR cross-contrast prediction, some prior work has attempted to combine it with multi-contrast reconstruction. A naive form of joint image synthesis and reconstruction, e.g., PROSIT \citep{matterncontrast} and rsGAN \citep{dar_prior-guided_2020}, involves generating a synthetic image from the reference scan via deterministic image translation and using it in a classical L2-regularized least-squares reconstruction. More recent work by \citet{xuan2022multimodal} proposed a joint image translation and reconstruction method which additionally accounts for misalignment between the reference and reconstructing images, with a follow-up work leveraging optimal transport theory \citep{wang2024spatial}. \citet{levac2023mri} (whose approach we refer to as "PnP-Diffusion") formulate guided reconstruction as a highly general Bayesian maximum \textit{a posteriori} estimation problem and solve it iteratively via Langevin update steps, using an image-domain diffusion model as the score function of the prior distribution. 
While these are promising directions, we propose an alternative approach that decomposes the multi-contrast problem in a more intuitive way: first, into two sub-problems, namely image modeling and image reconstruction; and second, within the image-domain model, the multi-contrast information is decomposed into content and style. This two-level decomposition results in a highly modular reconstruction algorithm with a built-in explanatory framework where (a) the guidance mechanism is a simple content-replacement operation, (b) the discrepancy between the supplied reference content and the true content of the target image represents a meaningful error term which can be minimized, and (c) the optimal content-encoding capacity of the model for a given two-contrast image dataset indicates the amount of shared structure that is available to be learned in this data and utilized in the reconstruction task.

\section{Methods}

\noindent
In this section, we describe our methodology, starting with the formulation of the reconstruction problem and introducing the general plug-and-play method for solving it. We first sketch the outline of our content consistency operator in Section \ref{sec:methods-problem}, followed by defining in Section \ref{sec:methods-model} the content/style model that is at the core of this operator, and finally constructing the operator itself and developing it into the full PnP-CoSMo iterative reconstruction algorithm in Section \ref{sec:methods-recon}.

\subsection{Reconstruction problem}
\label{sec:methods-problem}

\noindent
Given a set of $P$ acquired k-space samples $y \in \mathbb{C}^P$ and the MRI forward operator $A \in \mathbb{C}^{P \times Q}$, CS reconstruction of the image $x \in \mathbb{C}^Q$ with $Q$ voxels is given as
\begin{equation}
    \min_x || Ax - y ||_2^2 + \lambda || \Psi x ||_1,
    \label{eq:cs}
\end{equation}
where $\Psi$ is some sparsifying transform (e.g., wavelet) and $\lambda$ is the regularization strength. A commonly used algorithm to solve this optimization problem is ISTA, which iteratively applies the following two update steps:
\begin{equation}
    r^k \leftarrow \Psi^H \mathrm{soft}(\Psi x^{k-1}; \lambda),
    \label{eq:softthresh}
\end{equation}
\begin{equation}
    x^k \gets r^k - \eta A^H(Ar^k - y).
    \label{eq:dc}
\end{equation}
Eq. \eqref{eq:softthresh} performs soft-thresholding in the transform domain, thereby reducing the incoherent undersampling artifacts in image $x^{k-1}$, whereas \eqref{eq:dc} enforces soft data consistency on image $r^k$ by taking a single gradient descent step over the least-squares term, controlled by step size $\eta$.

Plug-and-play methods replace the analytical operation of \eqref{eq:softthresh} with off-the-shelf denoisers. A CNN-based denoiser is of special interest as it incorporates a learning-based component into iterative reconstruction. Given a CNN model $f_{\theta}$ trained to remove i.i.d. Gaussian noise from an image, PnP-CNN \citep{ahmad_plug-and-play_2020} modifies Eq. \eqref{eq:softthresh} to
\begin{equation}
    z^k \leftarrow f_{\theta}(x^{k-1}).
    \label{eq:pnpcnn}
\end{equation}

In guided reconstruction, a spatially aligned reference $x^\mathrm{ref}_1$ is available, which captures the same underlying \textit{semantic content} as the target reconstruction. Inspired by the PnP design, we cast the problem of incorporating prior information from $x^\mathrm{ref}_1$ into the reconstruction iterate $x_2^{k-1}$ as enforcing a hard consistency between this image iterate and its semantic content estimated from the reference. We propose a \textit{content consistency operator} $g_M(\cdot;c)$ such that
\begin{equation}
    z_2^k \leftarrow g_M(x_2^{k-1}; c),
    \label{eq:munit_op1}
\end{equation}
where a content/style model $M$ decomposes $x_2^{k-1}$ into content and style, followed by a replacement of this corrupted content with high-quality content $c$ derived from $x^\mathrm{ref}_1$ and composing the improved image $z_2^k$ from it. The model $M$ is learned separately via an offline procedure and constitutes the core of our plug-and-play design. Before formally defining the operator $g_M(.;c)$ and developing the reconstruction algorithm, we discuss in the following section the design and learning of the content/style model $M$.

\subsection{Content/Style modeling}
\label{sec:methods-model}
\noindent
Given an image dataset of two MR contrasts, we make the general assumption that there exists an underlying contrast-independent structure which, influenced by arbitrary contrast-specific factors, is rendered as the contrast images.
We formulate our content/style model based on the MUNIT framework \citep{huang2018multimodal}. We first define the model and its pre-training strategy in Section \ref{sec:methods-model-pretraining}, subsequently proposing a fine-tuning phase in Section \ref{sec:methods-model-pft} to adapt it to our reconstruction task. Finally, in Section \ref{sec:methods-model-localitybias}, we introduce the concept of \textit{content capacity} as a function of the model's architecture, which would directly influence the effectiveness of our content consistency operator developed using this model.

\subsubsection{Unpaired MUNIT pre-training}
\label{sec:methods-model-pretraining}
\noindent
We define an image domain as the set $\mathcal{X}_i$ of images of a given contrast comprising the dataset, where $i \in \{1,2\}$. "Content" $c \in \mathcal{C}$ is defined as the underlying contrast-independent structure and represented as a set of feature maps, whereas "style" $s_i \in \mathcal{S}_i$ corresponds to the various modes of variability in one domain which cannot be explained by the other, e.g., global effects of acquisition settings, contrast-specific tissue features, etc., and is represented as a low-dimensional vector. 

MUNIT posits the existence of functions $G_i^*: \mathcal{C}\times\mathcal{S}_i  \rightarrow \mathcal{X}_i$ and their inverses $E_i^* = (G_i^*)^{-1}$, and learns them jointly via unpaired training, given samples from marginal distributions $p(x_i)$. In practice, the encoder $E_i$ is split into the content encoder $E_i^c$ and the style encoder $E_i^s$. Thus, the content/style model is specified as $M = \{E_1^c, E_2^c, E_1^s, E_2^s, G_1, G_2\}$. The MUNIT loss function comprises four terms:
\begin{equation}
    \mathcal{L_\mathrm{MUNIT}} = \mathcal{L}_\mathrm{GAN} + 
    \alpha_1 \mathcal{L}_\mathrm{image}^\mathrm{self} + 
    \alpha_2 \mathcal{L}_\mathrm{content}^\mathrm{self} + 
    \alpha_3 \mathcal{L}_\mathrm{style}^\mathrm{self},
    \label{eq:munit_loss}
\end{equation}
where 
\begin{equation}
\begin{split}
\mathcal{L}_\mathrm{GAN} = \:&\mathbb{E}_{x_1\sim p(x_1), s_2\sim q(s_2)}[(1 - D_2(G_2(E_1^c(x_1), s_2)))^2] + \\ &\mathbb{E}_{x_2\sim p(x_2), s_1\sim q(s_1)}[(1 - D_1(G_1(E_2^c(x_2), s_1)))^2],
\end{split}
\end{equation}
\begin{equation}
\begin{split}
\mathcal{L}_\mathrm{image}^\mathrm{self} = \:&\mathbb{E}_{x_1\sim p(x_1)}[||x_1 - G_1(E_1^c(x_1), E_1^s(x_1))||_1] + \\ &\mathbb{E}_{x_2\sim p(x_2)}[||x_2 - G_2(E_2^c(x_2), E_2^s(x_2))||_1],
\end{split}
\end{equation}
\begin{equation}
\begin{split}
\mathcal{L}_\mathrm{content}^\mathrm{self} = \:&\mathbb{E}_{c_1\sim p(c_1), s_2\sim q(s_2)}[||c_1 - E_2^c(G_2(c_1, s_2))||_1] + \\ &\mathbb{E}_{c_2\sim p(c_2), s_1\sim q(s_1)}[||c_2 - E_1^c(G_1(c_2, s_1))||_1],
\end{split}
\end{equation}
\begin{equation}
\begin{split}
\mathcal{L}_\mathrm{style}^\mathrm{self} = \:&\mathbb{E}_{c_1\sim p(c_1), s_2\sim q(s_2)}[||s_2 - E_2^s(G_2(c_1, s_2))||_1] + \\ &\mathbb{E}_{c_2\sim p(c_2), s_1\sim q(s_1)}[||s_1 - E_1^s(G_1(c_2, s_1))||_1].
\end{split}
\end{equation}
$\mathcal{L}_\mathrm{image}^\mathrm{self}$ is the image recovery loss, which promotes preservation of the image information in the latent space. $\mathcal{L}_\mathrm{content}^\mathrm{self}$ and $\mathcal{L}_\mathrm{style}^\mathrm{self}$ are the content and style recovery losses, respectively, which specialize parts of the latent code into content and style components. $\mathcal{L}_\mathrm{GAN}$ is the adversarial loss, which enables unpaired training by enforcing distribution-level similarity between synthetic and real images via discriminators $D_1$ and $D_2$. $\alpha_1$, $\alpha_2$, and $\alpha_3$ are tuneable hyperparameters. In the present work, we consider the MUNIT training process as the pre-training stage of our content/style model.

\subsubsection{Paired fine-tuning}
\label{sec:methods-model-pft}
\begin{figure*}[tb!]
    \centering
    \includegraphics[width=0.95\textwidth]{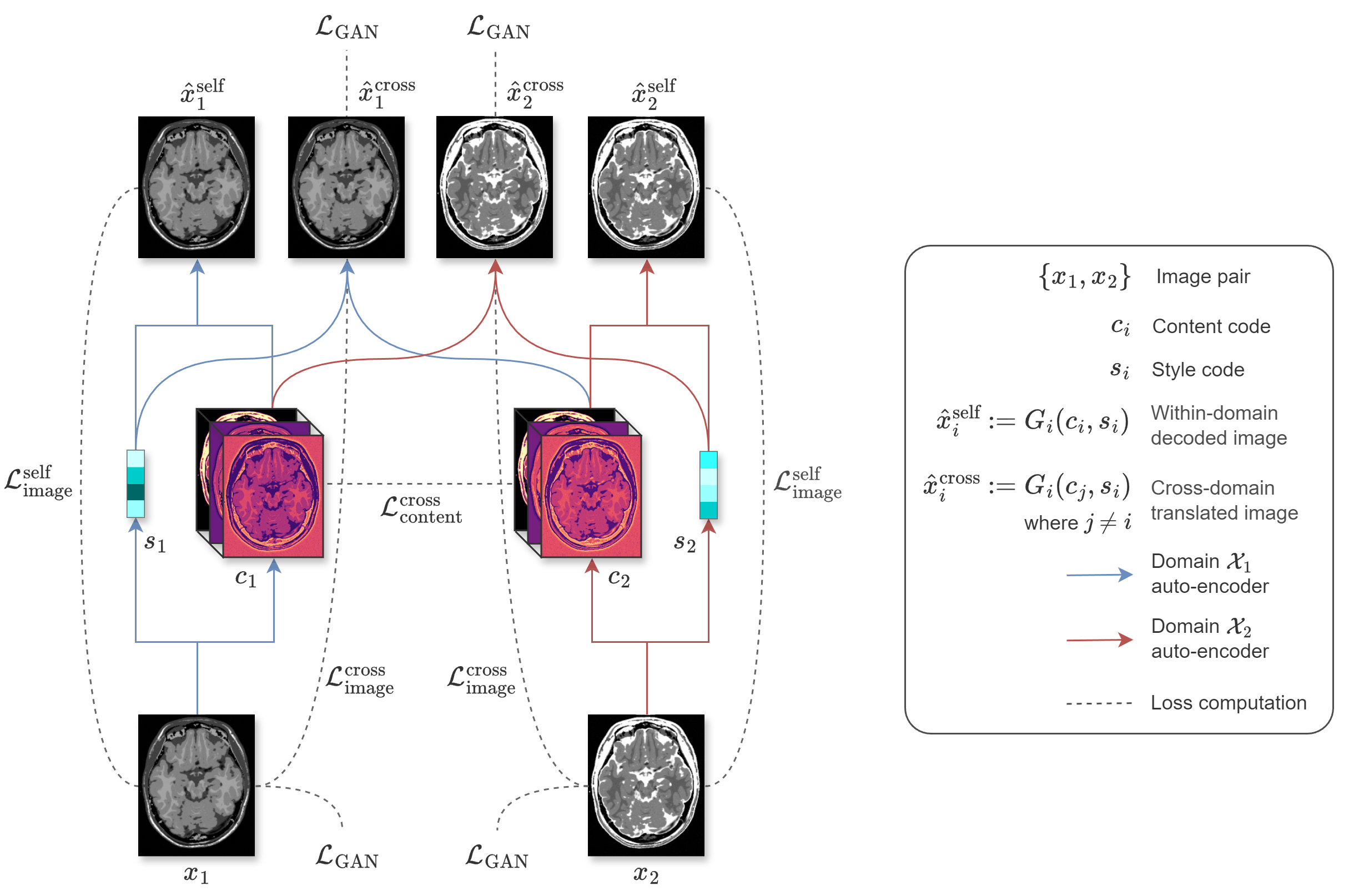}
    \caption{Our proposed paired fine-tuning (PFT) stage for improving the alignment of the content representations of the two domains given a small dataset of paired images.}
    \label{fig:pft}
\end{figure*}
\noindent
While the pre-training stage constitutes the bulk of the training and is responsible for learning the disentangled content/style representations, we propose an additional training phase for further improving the model's content preservation property. To this end, we introduce the paired fine-tuning (PFT) stage, which strongly aligns the content representations of the two domains given a limited amount of paired, spatially-aligned image data. Given samples from the joint distribution $p(x_1, x_2)$, our fine-tuning objective is expressed as
\begin{equation}
    \mathcal{L}_\mathrm{PFT} = \mathcal{L}_\mathrm{GAN} + 
    \beta_1 \mathcal{L}_\mathrm{image}^\mathrm{self} + 
    \beta_2 \mathcal{L}_\mathrm{image}^\mathrm{cross} + 
    \beta_3 \mathcal{L}_\mathrm{content}^\mathrm{cross},
\end{equation}
where 
\begin{equation}
\begin{split}
\mathcal{L}_\mathrm{image}^\mathrm{cross} = \mathbb{E}_{\{x_1, x_2\}\sim p(x_1, x_2)}[ \:&||x_2 - G_2(E_1^c(x_1), E_2^s(x_2))||_1 + \\ &||x_1 - G_1(E_2^c(x_2), E_1^s(x_1))||_1],
\end{split}
\end{equation}
\begin{equation}
\begin{split}
\mathcal{L}_\mathrm{content}^\mathrm{cross} = \mathbb{E}_{\{x_1, x_2\}\sim p(x_1, x_2)}[||E_1^c(x_1) - E_2^c(x_2)||_1].
\end{split}
\end{equation}
Among the newly introduced PFT loss terms here, $\mathcal{L}_\mathrm{image}^\mathrm{cross}$ is a pixel-wise image translation loss, which provides bidirectional image-level supervision, and $\mathcal{L}_\mathrm{content}^\mathrm{cross}$ is a paired content loss, which penalizes discrepancy between the content representations of the paired images. $\beta_1$, $\beta_2$, and $\beta_3$ are an additional set of hyperparameters. 
Note that our PFT preserves the bi-directionality of the content/style model, a property which will later be essential for generalization across different reconstruction tasks.

\subsubsection{Network architecture and content capacity}
\label{sec:methods-model-localitybias}

\noindent
Following the original MUNIT paper \citep{huang2018multimodal}, our content encoders $E_i^c$ consist of an input convolutional layer potentially followed by strided downsampling convolutions, and finally a series of residual blocks. Style encoders $E_i^s$ consist of input and downsampling convolutions followed by adaptive average pooling and a fully-connected layer that outputs the latent vector. Decoders $G_i$ follow a similar structure to the content encoders, except in reverse. Style is introduced into the decoder via AdaIN operations, which modulate the activation maps derived from the content.
We observe that in this architecture, the ratio between the content resolution and the image resolution reflects the level of local structure one expects to be shared between the two domains. This (relative) content resolution is thus an inductive bias built into the model, which we refer to as the model's \textit{content capacity} and define as
\begin{equation}
    J_M := \frac{H_c W_c}{H_x W_x},
    \label{eq:ccapacity}
\end{equation}
where $H_x$ and $W_x$ are the height and width of the input image and $H_c$ ($\leq H_x$) and $W_c$ ($\leq W_x$) are those of the corresponding content maps. This ratio is controlled by changing the number of up/downsampling blocks in the networks.
Our notion of content capacity is closely related to the more general concept of the locality bias of image-to-image models \citep{richardson_surprising_2021}. A content/style model with a high content capacity has a large content resolution, allowing it to learn a rich, highly expressive content representation that strongly influences the output's structure. Consequently, the model's style spaces are restricted to learn mostly low-level global features. By finding the \textit{optimal content capacity} for a given two-contrast image-domain dataset, one can quantify the amount of shared spatial information in this dataset, which will have consequences on the effectiveness of guidance in the reconstruction.

\begin{figure*}
    \centering
    \includegraphics[width=1\textwidth]{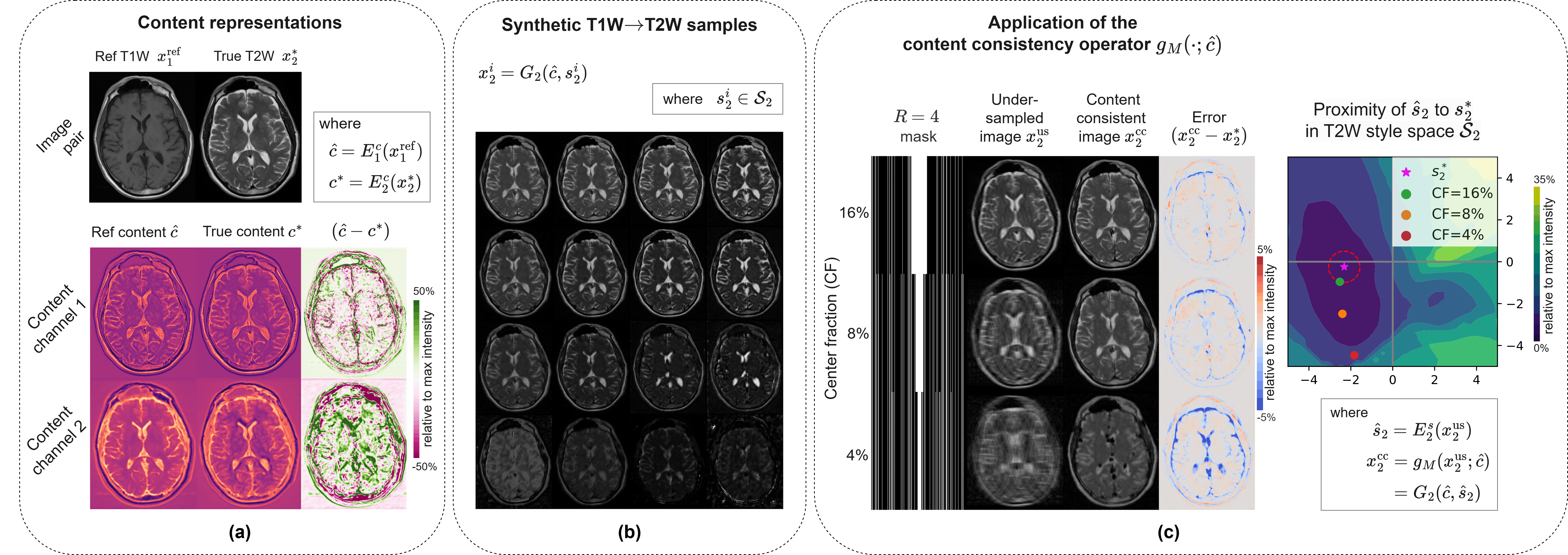}
    \caption{Demonstration of the content consistency operator $g_M(\cdot;c)$ based on a content/style model $M$. The image pair shown here is an example from the NYU fastMRI brain DICOMs, practical details regarding which are provided in Section \ref{expt_design}. (a) A T1W/T2W image pair and the corresponding content maps show that the two contents generally agree, while there is also some discrepancy between them. (b) Synthetic T2W images generated from 16 style codes grid-sampled from $\mathcal{S}_2$, showing that style encodes mostly the image contrast. (c) Ground truth T2W $x_2^*$ is corrupted to $x_2^\text{us}$ using $R$=4 undersampling with three different center fractions and the operator $g_M(\cdot;c)$ is applied to each case, showing that sampling more low-frequency lines leads to more accurate style estimate $\hat{s}_2$ and hence a better content-consistent image $x_2^\mathrm{cc}$. Contours in $\mathcal{S}_2$ indicate the level set of MAE of the synthetic images corresponding to styles from those regions. The red circle indicates a style estimation NMSE of 0.1.}
    \label{fig:munit_op}
\end{figure*}

\subsection{Iterative reconstruction based on content consistency}
\label{sec:methods-recon}
\noindent
Given the content/style model $M$, we consider $\mathcal{X}_1$ and $\mathcal{X}_2$ as reference and target domains, respectively, in our guided reconstruction task, and define the content consistency operator $g_M(\cdot;c)$ from Eq. \eqref{eq:munit_op1} as
\begin{equation}
    x^\mathrm{cc}_2 = g_M(x^\mathrm{us}_2; \hat{c}) :=  G_2(\hat{c}, E^s_2(x^\mathrm{us}_2)),
    \label{eq:munit_op2}
\end{equation}
where $x^\mathrm{us}_2$ is the image containing undersampling artifacts, $x^\mathrm{cc}_2$ is the content-consistent image, and $\hat{c} = E^c_1(x^\mathrm{ref}_1)$ is the reference content. This operation improves $x^\mathrm{us}_2$ by simply replacing its aliased content with the content estimated from $x^\mathrm{ref}_1$, a rule which will later be softened with \eqref{eq:content_update}.

Note that this is a radical operation as it discards all structure contained in $x^\mathrm{us}_2$, retaining only a compact style code $\hat{s}_2 = E_2^s(x^\mathrm{us}_2)$. Let $x_2^*$ be the ground truth reconstruction with content $c^*$ and style $s_2^*$. 
The success of our content consistency operation depends on three conditions: (a) $\hat{c}$ and $c^*$ are highly descriptive representations of the underlying anatomy, (b) $\hat{c}$ is close to $c^*$, and (c) $\hat{s}_2$ is close to $s_2^*$. 
The first condition is satisfied provided that there is a sufficient amount of shared structure between the reference and target domains and that the model has a high content capacity $J_M$ for capturing highly expressive content. The second condition is roughly satisfied, as seen in Fig. \ref{fig:munit_op}a, because the model is explicitly trained to minimize content error. This point will be revisited shortly in the coming paragraphs. On the other hand, it is not self-evident that the third condition should hold too, and it hence deserves a closer examination. 
If the first condition is satisfied, the optimal model will have a high content capacity and consequently, the style variable would be confined to representing low-level global image features, e.g., contrast variations (see Section \ref{sec:methods-model-localitybias}). This can be observed empirically, as shown in Fig. \ref{fig:munit_op}b. It is a well-known fact that image contrast is contained prominently in the center of the k-space. Hence, the estimate $\hat{s}_2$ can be made arbitrarily close to $s_2^*$ by sufficiently sampling the k-space center, as demonstrated empirically in Fig. \ref{fig:munit_op}c.

Applying data consistency update \eqref{eq:dc} following $g_M(\cdot;\hat{c})$ in repetition yields an ISTA-based iterative scheme. Here, our content consistency update complements the data consistency update in the sense that while the latter forces the image estimate to be consistent with the given (measured) k-space data $y$, the former forces it to be consistent with the given (prior) content $\hat{c}$.

\begin{algorithm}[h!]
    \caption{PnP-CoSMo iterative guided reconstruction}
    \label{alg:pnp_munit}
    \begin{algorithmic}[1]
        \Require $y$, $A$, $x^\mathrm{ref}_1$, $M$, $\eta$, $\gamma$  
        \Comment{where $M=\{E_1^c,E_1^s,E_2^c,E_2^s,G_1,G_2\}$}
        \State $k \gets 0$
        \State $x_2^k \gets A^H y$                     \Comment{Initialize reconstruction}
        \State $c^k \gets E_1^c(x^\mathrm{ref}_1)$              \Comment{Initialize content}
        \Repeat
        \State $k \gets k + 1$ 
        \State $z_2^k \gets g_M(x_2^{k-1}; c^{k-1})$  \Comment{Content consistency update}
        \State $x_2^k \gets z_2^k - \eta A^H(Az_2^k - y)$  \Comment{Data consistency update}
        \State $c^k\gets c^{k-1} - \gamma \nabla_c || A G_2(c^{k-1}, E_2^s(x_2^k)) - y ||_2^2 $  \Comment{Content refinement (CR) update} 
        \Until{convergence} \\
        \Return $x_2^k$
    \end{algorithmic}
\end{algorithm}

The core assumption of an idealized content/style model is that contents $\hat{c}$ and $c^*$ are identical. However, in reality, this assumption will not hold, and a discrepancy between the two contents is to be expected. There are two possible sources of this discrepancy: (a) model-related, e.g., fundamental limits such as irreducible error\footnote{If perfectly zero error was possible, one could perfectly predict the (distribution of) target image from the reference without any measurement.} and practical issues like sub-optimal network design and training, and (b) reference image-related, e.g., presence of artifacts independent of the target reconstruction. We refer to this gap between $\hat{c}$ and $c^*$ as \textit{content discrepancy}. Fig.\ref{fig:munit_op}a shows content discrepancy as observed empirically.
During reconstruction, the error in $\hat{c}$ would limit the efficacy of the content consistency operator $g_M(\cdot;\hat{c})$, affecting the reconstruction quality. While model-related discrepancy is partly tackled by PFT (Section \ref{sec:methods-model-pft}), we now propose a \textit{content refinement} (CR) procedure to correct for the remaining discrepancy in the reconstruction stage. 
Since we have no direct access to the true content $c^*$, but only to undersampled k-space measurements $y$, we aim at solving the following minimization sub-problem in CR
\begin{equation}
    \min_c || A G_2(c, \hat{s}_2) - y ||_2^2,
    \label{eq:content_opt}
\end{equation}
starting from the initial point $\hat{c}=E^c_1(x^\mathrm{ref}_1)$ and for a given style estimate $\hat{s}_2$. The augmented forward operator $AG_2(\cdot)$ is a composition of the linear MRI forward operator $A$ and the non-linear content/style decoder $G_2$, and it maps the content domain to the k-space domain. The error between the predicted k-space $AG_2(c,\hat{s}_2)$ and the measured data $y$ serves as a proxy for content discrepancy, which can be computed and minimized during the reconstruction. We approximate the solution with a single gradient descent step
\begin{equation}
    c^{k} \gets c^{k-1} - \gamma \nabla_c || A G_2(c^{k-1}, \hat{s}_2^k) - y ||_2^2,
    \label{eq:content_update}
\end{equation}
initialized as $c^0 \gets E^c_1(x^\mathrm{ref}_1)$ and updated every $k^\text{th}$ iteration with step size $\gamma$ following the data consistency and content consistency updates.
Hence, by aligning the content $c$ with k-space data $y$ and correcting the discrepancy, the CR module aligns content consistency updates with data consistency updates. 
With this additional component in place, we obtain our PnP-CoSMo reconstruction algorithm (Algorithm \ref{alg:pnp_munit}). 
Note that, unlike content, the style of the image estimate need not be explicitly aligned with the k-space. Since the style information is assumed to be derived predominantly from the k-space data, the data consistency update applied to the image implicitly "corrects" its style. Thus, as the image converges, we hypothesize that style would converge as well.

\section{Experimental setup}
\label{expt_design}

\noindent
In order to comprehensively evaluate PnP-CoSMo, we conducted four sets of experiments, each addressing different questions. First, a set of detailed analyses of our method using simulated MRI images tested the influence of various hyperparameters and of structural data errors on the reconstruction quality as well as the algorithm's convergence properties. Second, with the aim of evaluating against a broad set of state-of-the-art baselines and multiple reconstruction tasks, we benchmarked PnP-CoSMo on the public NYU fastMRI brain data \citep{knoll2020fastmri,zbontar1811fastmri}. Third, we performed a benchmark and ablation study on two challenging clinical multi-coil datasets with real-world data constraints, following which a small-scale radiological evaluation was conducted to obtain qualitative radiological feedback on a small data sample.
Without loss of generality, we considered the case of reconstructing T2W scans using T1W references and focused on head applications, except in the NYU benchmark, where the reverse case of T2W-guided T1W reconstruction was considered as well. Supplementary results on the NYU dataset involving the two additional contrasts, namely FLAIR and gadolinium-enhanced T1W (T1WGd), are provided in \ref{appendix:nyu_benchmark_extended}.

\subsection{Analyses on simulated MR data}
\noindent
The goal of our simulation experiments was to explore the hyperparameters and empirically test several properties of our algorithm under a tightly controlled setting. 
To this end, we used simulated T1W and T2W images based on BrainWeb phantoms \citep{collins1998design}. BrainWeb provides 20 anatomical models of the normal brain, each comprised of fuzzy segmentation maps of 12 tissue types. The 20 volumes were first split into the model training (18), model validation (1), and reconstruction testing (1) subsets. 2D T1W and T2W spin-echo scans in the transversal plane were simulated using TE/TR values randomly sampled from realistic ranges, specifically TR $\in[300,700]$ and TE $\in[5,25]$ for T1W scans, and TR $\in[1800,2200]$ and TE $\in[70,120]$ for T2W scans. For testing the reconstruction, 2D single-coil T2W k-space data were simulated via Fourier transform and 1D Cartesian random sampling at various accelerations. 

Among the various hyperparameters of our framework, the most influential are the content capacity $J_M$ of the content/style model (Eq. \ref{eq:ccapacity}), the disentanglement loss weights $\alpha_2$ and $\alpha_3$ of the model pre-training (Eq. \ref{eq:munit_loss}), and the CR step size $\gamma$ of the PnP-CoSMo algorithm (Eq. \ref{eq:content_update}). 
While $J_M$ defines the richness of the contrast-independent content the model is capable of learning, $\alpha_2$ and $\alpha_3$ determine the degree of disentanglement of the content and style representations the model achieves during the bulk of its training. Both are necessary preconditions for our content consistency operator, and hence for PnP-CoSMo, to be effective. In PnP-CoSMo, the CR step size $\gamma$ controls the algorithm's efficacy in correcting errors in the reference content, thereby directly influencing reconstruction quality. Unlike other hyperparameters, $\gamma$ is specific to the reconstruction problem and depends on the reconstruction setup, e.g., k-space undersampling.
We thus analyzed the effects of these specific hyperparameters using the BrainWeb dataset, obtaining insights on their optimal settings,  while additionally analyzing other aspects such as algorithm convergence and the robustness of the CR module.

\subsection{Benchmark on public NYU fastMRI DICOM data}
\label{expt:nyu}
\noindent
In the NYU benchmark, we compared PnP-CoSMo against state-of-the-art reconstruction methods belonging to both end-to-end and plug-and-play paradigms.
First, we sought to answer the following question: Can PnP-CoSMo, which requires only image-domain training data, outperform end-to-end methods, which additionally require k-space training data and are optimized to directly solve the reconstruction problem? 
Furthermore, end-to-end methods are task-specific models where the reference and target contrasts are fixed, e.g., T1W as reference and T2W as target. On the other hand, the content/style model used in PnP-CoSMo does not possess this kind of unidirectionality. Hence, we sought to assess the generalizability of PnP-CoSMo across two different reconstruction tasks: (a) T2W reconstruction with guidance from T1W reference and (b) T1W reconstruction with guidance from T2W reference.
Finally, we compared PnP-CoSMo against other plug-and-play methods, which share a similar design pattern as ours, where the modeling/learning problem is decoupled from the reconstruction problem, and the requirement is only image-domain training data.

The public NYU fastMRI brain dataset includes multi-coil raw data and DICOM scans of four contrasts: T1W with and without gadolinium agent, T2W, and FLAIR. 
For multi-contrast experiments, however, the NYU raw dataset does not provide subject labels for pairing the contrasts. Therefore, instead of the NYU raw data, we used the NYU DICOMs in this study, following \citet{xuan2022multimodal}. A paired DICOM subset of 327 subjects was obtained based on T2W and non-gadolinium T1W scans, all of them being either 2D spin echo or turbo spin echo acquisitions in the transverse plane with median in-plane FOV of 220$\times$220 mm$^2$ for both, in-plane voxel size of 0.69$\times$0.69 mm$^2$ for T1W scans and 0.57$\times$0.57 mm$^2$ for T2W scans, slice thickness of 5 mm for both, FA=20-150$^{\circ}$, TR=220-874 ms, TE=2.46-17 ms, and ETL=1-5 for T1W scans, and FA=90-180$^{\circ}$, TR=2696-8910 ms, TE=9.9-119 ms, and ETL=11-27 for T2W scans.
All T1W scans were rigidly registered with the corresponding T2W scans. Single-coil k-space data was simulated via Fourier transform with 1D Cartesian random sampling at $R\in \{2,3,4,5\}$ and added Gaussian noise of $\sigma = 0.01 \mathrm{max}(x_2^*)$. 
 
The dataset was split into three folds at the subject-level for cross-validation based on random shuffling, with two partitions representing training data and one representing the test data in each fold. For PnP-CoSMo, all images from the training set were used to pre-train (ignoring the pairing information) and fine-tune (utilizing the pairing information) the content/style model, whereas the end-to-end methods were trained on paired examples composed of the reference image, the target contrast k-space, and its ground truth image. The plug-and-play baselines were trained, like PnP-CoSMo, on the full image-domain data in the cross-validation training folds.
We benchmarked PnP-CoSMo against three end-to-end methods and three plug-and-play methods. Among the end-to-end baselines we included one well-known single-contrast network, namely MoDL \citep{aggarwal2018modl}, and two recent multi-contrast networks, namely MTrans \citep{feng2022multimodal} and MC-VarNet \citep{lei2023decomposition}. These networks were evaluated on the aforementioned two reconstruction tasks, assessing both their in-distribution and out-of-distribution performance. Specifically, for the T2W reconstruction task, "in-distribution" means that the end-to-end model was trained to reconstruct T2W images, whereas "out-of-distribution" means the model was trained to reconstruct T1W images but applied on the T2W reconstruction task, and \textit{vice versa}. While MC-VarNet and MTrans used the reference image information,  MoDL ignored the reference information. Among the plug-and-play baselines we included a single-contrast method, namely PnP-CNN \citep{ahmad_plug-and-play_2020}, and two guided methods, namely PnP-Diffusion \citep{levac2023mri} and PROSIT \citep{matterncontrast}. Note that these methods, like PnP-CoSMo, are not contrast-task-specific in the same way end-to-end models are, and thus the concept of in-distribution and out-of-distribution tasks does not apply to them. 

\begin{table}
    \small
    \centering
    \caption{Overview of the sequences in the LUMC datasets.}
    \begin{tabular}{lllll}
    \hline
 \multirow{2}{*}{\vtop{\hbox{\strut Sequence} \hbox{\strut Params}}}& \multicolumn{2}{c}{LUMC-TRA }& \multicolumn{2}{c}{LUMC-COR} \\
 \cline{2-5}
 &  3D T1W TFE&  2D T2W TSE&  3D T1W TSE& 2D T2W TSE\\
 \hline
         FA (deg)&  8 &  90 &  80-90& 90 \\
         TR (ms)&  9.8-9.9 &  4000-5000 &  500-800& 2000-3500 \\
         TE (ms)&  4.6 &  80-100 &  6.5-16& 90-100\\
         ETL&  200 &  14-18 &  10-13& 17-19 \\
         \vtop{\hbox{\strut Voxel size} \hbox{\strut(mm)}} & \vtop{\hbox{\strut 0.98$\times$0.99} \hbox{\strut $\times$0.91}}& 0.4$\times$0.54 & \vtop{\hbox{\strut 0.59$\times$0.62} \hbox{\strut $\times$1.19}}  &  0.39$\times$0.47 \\
 \vtop{\hbox{\strut Slice thick.} \hbox{\strut(mm)}} & -- & 3 & -- &2 \\
         FOV (mm) & \vtop{\hbox{\strut 238$\times$191} \hbox{\strut $\times$218}} & 238$\times$190 & \vtop{\hbox{\strut 130$\times$238} \hbox{\strut $\times$38}} & 130$\times$197 \\
         Num slices & -- & 50& -- &15 \\
 \hline
    \end{tabular}
    \label{tab:lumc_protocols}
\end{table}

\begin{table}[tb!]
    \centering
    \caption{LUMC data split. The splits were made on the subject-level, i.e., each subject belonged to exactly one subset. Splits containing \textit{only} image-domain data (i.e., no raw k-space data), specifically \textit{image-train} and \textit{image-val}, are marked with $^\mathfrak{I}$. Thus, the training of only image-domain models is feasible, thereby representing a real-world problem where end-to-end methods are not feasible. The \textit{recon} split additionally included raw T2W multi-coil k-space and coil sensitivities, which were used for tuning and evaluating the reconstruction algorithms in a two-fold cross-validation setup. All data samples were used in the experiment.}
    \begin{tabular}{clccc}
     \hline
      Dataset&Split&  Subjects &Sessions & Scans (T1W / T2W)\\
      \hline
        \multirow{5}{*}{LUMC-TRA}&\textit{image-train}$^\mathfrak{I}$ &   295& 418& 360 / 415\\
        &\textit{image-val}$^\mathfrak{I}$ & 16& 17& 17 / 17\\
        &\textit{recon}  & 38& 52& 52 / 52\\
         & Total& 339& 487&429 / 484 \\
         \hline
         \multirow{5}{*}{LUMC-COR}& \textit{image-train}$^\mathfrak{I}$   & 242& 277& 269 / 272\\
         & \textit{image-val}$^\mathfrak{I}$    & 18& 18& 18 / 18\\
         & \textit{recon}   & 31& 35& 35 / 35\\
         & Total& 291& 330&322 / 325\\
         \hline
         \end{tabular}
    \label{tab:lumc_splits}
\end{table}

\subsection{Benchmark and ablation on in-house LUMC multi-coil datasets}
\noindent
In our benchmark on clinical multi-coil data, the goal was to test PnP-CoSMo on a constrained real-world problem where no raw k-space data was available for training, with only the image-domain training data at our disposal. This is often the case, as in clinical practice, raw data is discarded after acquisition, and only the final reconstructions are retained. Moreover, T1W/T2W images are often not fully subject-wise paired, e.g., due to missing acquisitions, representing a realistic case of data imbalance.

Our in-house data consisted of brain scans of patients from LUMC, the use of which was approved for research purposes by the institutional review board. A total of 1669 brain scans were obtained from 817 clinical MR examinations of 630 patients acquired on 3T Philips Ingenia scanners. We focused on accelerating two T2W sequences: (a) 2D T2W TSE transversal and (b) 2D T2W TSE coronal. For guidance, two corresponding T1W sequences were used: (a) 3D T1W TFE transversal and (b) 3D T1W TSE coronal. The transversal and coronal protocols were considered as two separate datasets, namely LUMC-TRA and LUMC-COR. Table \ref{tab:lumc_protocols} shows an overview of the four sequences. Note that the in-plane resolution of the T1W scans was 1.3-2.5 times as low as that of the T2W scans, making these datasets more challenging for T1W-guided T2W reconstruction.

Each of the two LUMC datasets was split into \textit{image-train}, \textit{image-val}, and \textit{recon}. The former two splits contained only image-domain data and no k-space raw data, and were used to train and validate the content/style models. The models were pre-trained on the full \textit{image-train} splits, ignoring any pairing between T1W and T2W scans. Additionally, 20 subjects from the \textit{image-train} split were designated for PFT, where the pairing information was used in training, and the reference scans were aligned via registration. The \textit{recon} split additionally included multi-coil T2W raw data, which comprised 6-channel (LUMC-TRA) and 13-channel (LUMC-COR) k-space and coil sensitivity maps. This k-space was already undersampled (1D Cartesian random) at acquisition with clinical acceleration of $R$=1.8-2. We further undersampled it retrospectively to higher accelerations $R\in\{4,6,8,10\}$ by dropping subsets of the acquired lines. 
The evaluation was performed on the \textit{recon} split using a two-fold cross-validation setup where the subjects in this split were divided into two folds. At a given moment, one fold was used to tune the reconstruction-time hyperparameters (such as CR step size $\gamma$ for PnP-CoSMo, regularization weights for classical baselines, etc.), whereas the other was used as a test set to compute the metrics, thus enabling a robust evaluation on all samples in the split.
An overview of the data split is shown in Table \ref{tab:lumc_splits}. Spatially aligned images required by all but the unpaired training set were obtained via rigid registration, and the reference T1W images were resampled to the T2W resolution. During unpaired training, all images were resampled to the median T2W resolution.

Given the absence of k-space training data, end-to-end reconstruction methods were not feasible as baselines. Hence, we compared PnP-CoSMo with only the feasible types of baselines, i.e., classical, semi-classical, and plug-and-play reconstruction and image-to-image translation. 
Among classical methods, we used the unguided $L_1$-wavelet CS (CS-WT) and the guided STV-based CS (CS-STV) \citep{ehrhardt_multicontrast_2016}. As an unguided plug-and-play baseline, we used PnP-CNN \citep{ahmad_plug-and-play_2020}, and as a representative image translation baseline, we compared against MUNIT itself. Using MUNIT, deterministic image translation was approximated by combining the reference content with 200 randomly sampled T2W style codes and taking a pixel-wise mean of the resulting synthetic images to obtain a single synthetic image. As a semi-classical guided baseline, we used PROSIT \citep{matterncontrast}, which combines deterministic image translation with L2-regularized least-squares reconstruction. 
Additionally, as an ablation study for PnP-CoSMo, we ablated PFT and CR to assess their contribution, and finally, as an upper bound for PnP-CoSMo representing zero content discrepancy, we disabled PFT and CR and used the ideal content $c^*$. In both NYU and LUMC benchmarks, we used three perceptual metrics for evaluation: SSIM, HaarPSI, and DISTS. While SSIM is used commonly, HaarPSI and DISTS are known to correlate better with visual judgment of image quality \citep{kastryulin_image_2023}. All three metrics are bounded in $[0,1]$ where $1$ represents perfect image quality. We conducted paired Wilcoxon signed-rank tests to measure statistical significance when comparing pairs of algorithms.

\subsection{Small-scale qualitative radiological evaluation}
\noindent
Finally, as a supplement to the LUMC benchmark, we conducted a preliminary and qualitative radiological evaluation with the goal of assessing the quality of PnP-CoSMo reconstructions in a concrete clinical context. The experiment was conducted on a small sample of the LUMC-TRA \textit{recon} set. First, all 20 subjects from this set (see Table \ref{tab:lumc_splits}) were screened by a junior radiologist for the presence of brain metastases, resulting in three scans included in this part of the experiment. 
The qualitative scoring was then performed independently by two senior neuroradiologists. The scoring system comprised two sets of criteria, namely visual quality and pathology. Four visual quality criteria were used, namely sharpness, noise, artifacts, and tissue contrast between gray and white matter and CSF. For pathology, we used three criteria, namely the number and sharpness of hyperintense areas within or surrounding metastases and the overall diagnostic quality of the scan for brain metastases. The images were scored using a five-point Likert scale \citep{mason2019comparison}: (1) non-diagnostic, (2) poor, (3) fair, (4) good, and (5) excellent diagnostic quality. 

Furthermore, we used PnP-CNN as a non-guided plug-and-play reconstruction baseline.
We evaluated the two methods at four clinically realistic accelerations of $R \in \{3,4,5,6\}$, which are close to the routine clinical acceleration of $R$=2 for this sequence. In total, nine images per patient, including the (2$\times$4=8) reconstructions and the clinical ground truth (acquired at $R$=2 and reconstructed by the vendor software), were presented to the two observers who scored each image individually, blinded to the reconstruction method and $R$. 
The scores were then processed as follows. Given a reconstruction method and $R$, the scores for each Likert-scale criterion were expressed as an aggregate by computing the mean across the three subjects and the two observers. The number of visible metastases was expressed as the fraction of total metastases across the three subjects. 
Moreover, the agreement between the two observers was measured using Spearman's rank correlation coefficient, computed separately on the Likert-scale scores and the metastasis count.
Note that these scores used in this small-scale evaluation are only meant to aid qualitative comparisons between the reconstruction algorithms and not to draw any strong quantitative inferences.

\subsection{Implementation details}
\noindent
We used a similar general residual architecture for content/style encoders and decoders as in the original MUNIT paper \citep{huang2018multimodal}, except with an additional layer at the end of the content encoder to produce content maps of the specified number of channels. We used 2 or 4 content channels, depending on the content downsampling factor. The discriminators were implemented as multi-scale PatchGAN networks, enabling them to locally assess the input images for realism at different scales. We used 1 scale in the BrainWeb simulations and three scales in the NYU and LUMC benchmark. We additionally conditioned the discriminators on foreground masks of the images to penalize the background signal, and we used spectral normalization \citep{miyato2018spectral} to stabilize the GAN training. 

Based on the hyperparameter analyses on BrainWeb synthetic data, viable ranges of values of the content capacity $J_M$, the disentanglement weights $\alpha_2$ and $\alpha_3$, and the step size $\gamma$ of the CR module were obtained (presented later in Section \ref{results:sim}) and set for the subsequent NYU and LUMC benchmarks. Specifically, $J_M$ was set depending on the resolution of the reference contrast of a given dataset, e.g., $J_M$=1 for the NYU data and $J_M=1/2^2$ for the LUMC data. We used $\alpha_2=\alpha_3=1$ as a reasonable choice for both the NYU and LUMC datasets. Finally, the CR step size was set to $\gamma=0.1$ for the NYU benchmark. In case of LUMC datasets, which consist of more complex multi-coil measurements, a search space was defined based on the BrainWeb results, and the optimal $\gamma$ was obtained by tuning the values per acceleration. The other hyperparameters, such as the image recovery loss weight $\alpha_1$ as well as the fine-tuning loss weights $\beta_1$, $\beta_2$, and $\beta_3$, were set to a simple and reasonable unit value consistently throughout. While additional hyperparameter optimizations may be possible, the above settings were sufficient to demonstrate the performance of PnP-CoSMo in comparison with strong baselines, fulfilling the aim of these benchmarks. In the synthetic BrainWeb experiments, all content/style models were pre-trained for 200k and were not fine-tuned for the sake of simplicity, unless stated otherwise in Section \ref{results:sim}. In both NYU and LUMC benchmarks, the content/style models were pre-trained for 400k iterations and fine-tuned for 50k iterations.  At reconstruction time, the number of PnP-CoSMo iterations was fixed to 200 consistently. 

All software was implemented in PyTorch, and image registration was performed using Elastix \citep{klein2009elastix}. All training runs were performed on a compute node with an NVIDIA Quadro RTX 6000 GPU.

\section{Results}
\label{results}

\noindent
The results of our extensive evaluation, as reported in the following sections, demonstrate the performance of PnP-CoSMo across synthetic and real-world MRI datasets, starting with detailed analyses on the BrainWeb synthetic data (Section \ref{results:sim}) that thoroughly examine PnP-CoSMo in terms of its most important hyperparameters and its robustness and convergence properties.
On the public NYU fastMRI benchmark (Section \ref{results:nyu_benchmark}), PnP-CoSMo is quantitatively compared against three end-to-end and three plug-and-play baselines, demonstrating superior quality and generalizability despite its lack of dependence on k-space training data. On in-house LUMC multi-coil raw data (Section \ref{results:lumc_benchmark}) --- constrained by partially paired image-domain training data and absent k-space training data --- PnP-CoSMo is compared against classical unguided and guided methods, plug-and-play approaches, and image-to-image translation techniques, confirming its superior performance on challenging real-world data, along with an ablation study verifying the contributions of the PFT and CR components. This is supplemented by a qualitative radiological evaluation (Section \ref{results:radeval}) providing initial clinical feedback.

\subsection{Analyses on simulated MR data}
\label{results:sim}

\noindent
In the following sections, we report our analyses on the synthetic BrainWeb dataset, exploring the influence of settings controlling both content/style modeling and iterative reconstruction stages of our framework. These are provided in Sections \ref{results:sim_cosmoprior_analyses} and  \ref{results:sim_cr_analyses}, respectively.
The most consequential hyperparameters of our framework are the content capacity $J_M$ and the disentanglement loss weights $\alpha_2$ and $\alpha_3$ of the content/style model, and the CR step size $\gamma$ of the PnP-CoSMo algorithm. 
Based on the sensitivity analyses presented here, we obtain transferrable insights into these settings.
Additionally, Section \ref{results:sim_cr_analyses} reports the effectiveness of the CR component in correcting contrast-specific structural misalignments, while also reporting a convergence analysis of PnP-CoSMo quantifying both its contribution and computational cost.

\begin{figure*}[h!]
    \centering
    \includegraphics[width=1\textwidth]{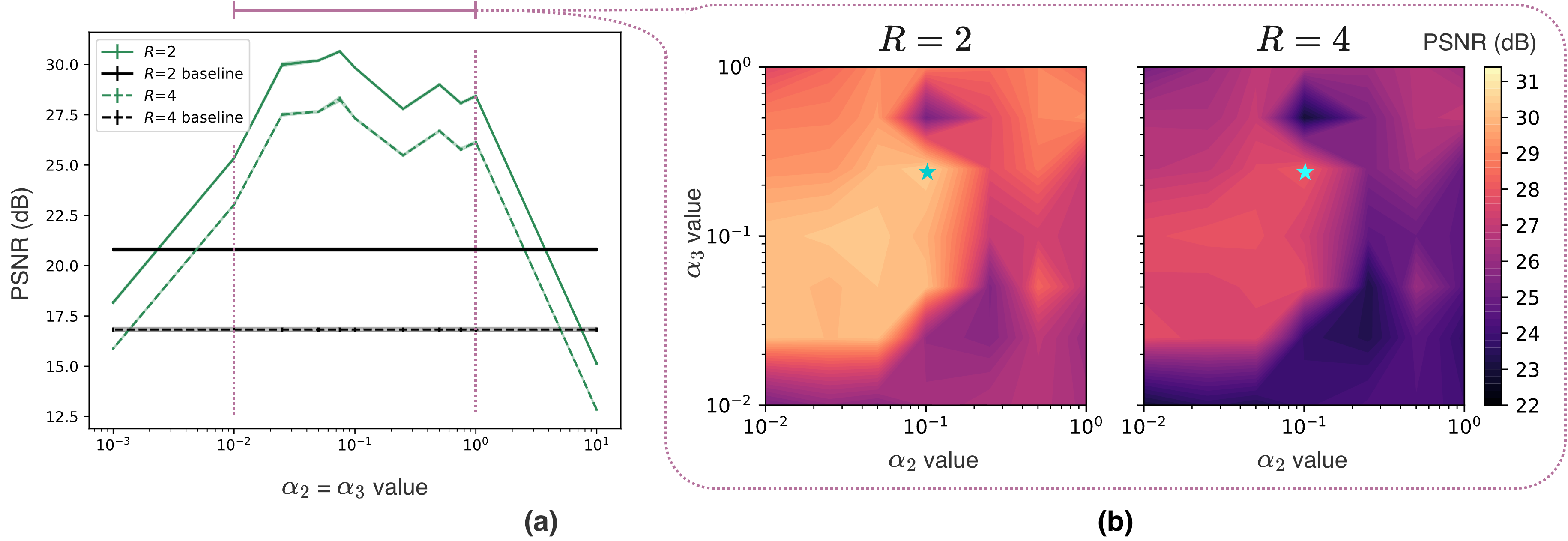}
    \caption{Sensitivity analysis for the disentanglement loss weights $\alpha_2$ and $\alpha_3$ on synthetic BrainWeb data. (a) The plot shows the effect of $\alpha_2$ and $\alpha_3$ on the reconstruction PSNR (dB) of PnP-CoSMo at accelerations $R\in \{2,4\}$ computed on BrainWeb test volume (300 slices). The black horizontal lines show the baseline reconstruction quality of $L_1$-wavelet compressed sensing (CS-WT). $\alpha_2 = \alpha_3 \in [0.01,1]$ is observed to be a reasonable range of settings. Outside of this range, the quality of PnP-CoSMo reconstructions sharply drop below the CS-WT baseline. A lower disentanglement strength results in insufficient separation of content from style, violating the modeling assumptions and resulting in worse reconstructions. Whereas, an excessive focus on disentanglement sacrifices preservation of image information in the latent space, leading to highly lossy encoding and decoding in the model and, hence, to worse reconstructions. 
    (b) Expanded analysis of the region $\alpha_2, \alpha_3 \in [0.01,1]$ of the parameter space, exploring cases where $\alpha_2 \neq \alpha_3$. An asymmetry in the sensitivity of PSNR is observed between $\alpha_2$ and $\alpha_3$, suggesting an unequal contribution to the level of content/style disentanglement in the model. This trend is consistent across acceleration factors, as seen across $R=2$ and $R=2$. The optimal setting is found to be $\alpha_2=0.1$ and $\alpha_3=0.25$ at both accelerations (annotated by cyan stars), with PSNR of 30.47 dB and 28.12 dB at $R=2$ and $R=4$, respectively.}
    \label{fig:sim_disent}
\end{figure*}

\begin{figure*}[h!]
    \centering
    \includegraphics[width=0.9\textwidth]{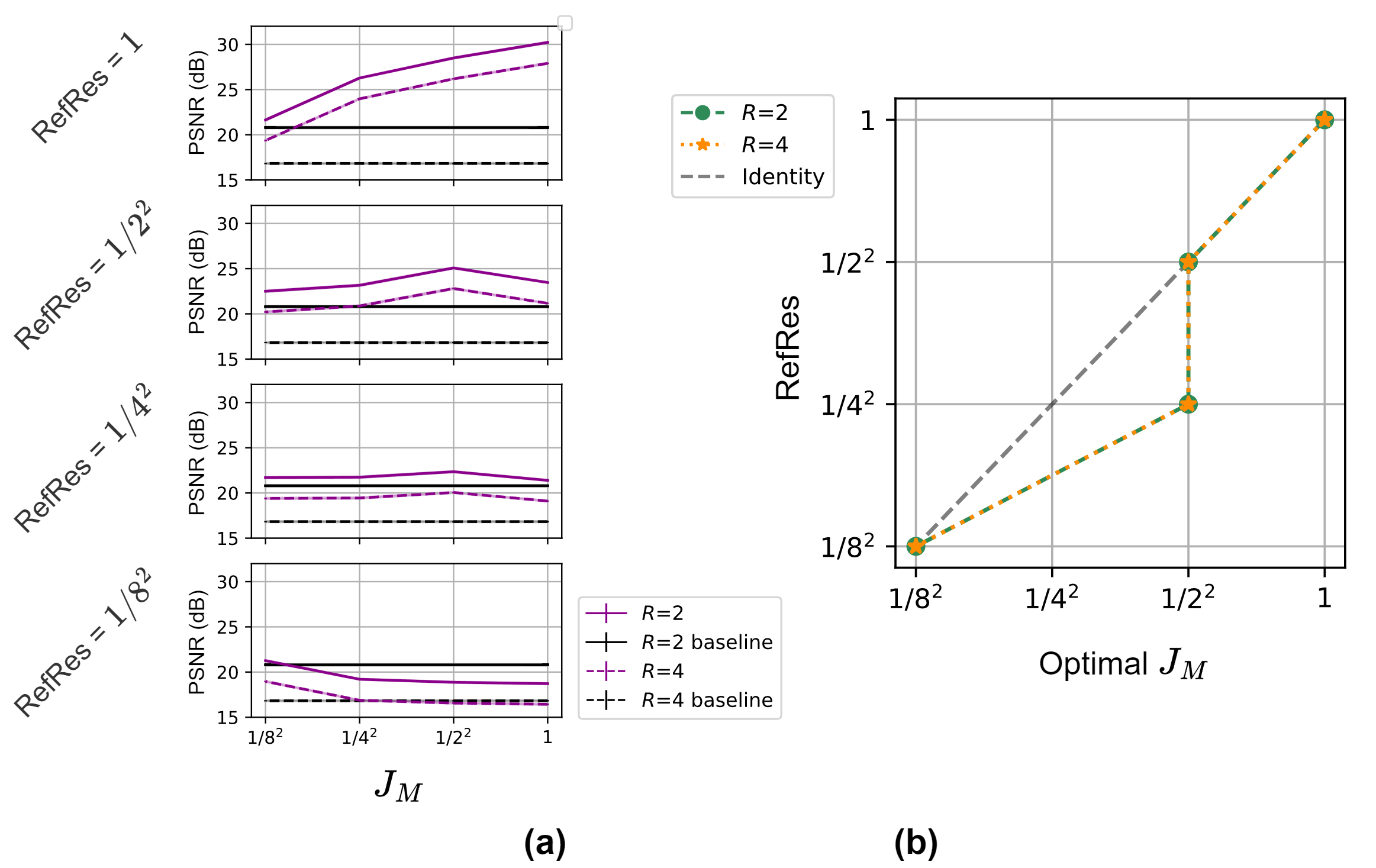}
    \caption{Analysis of the model's content capacity $J_M$ on synthetic BrainWeb data. The plots shows the dependence of the reconstruction quality on the reference resolution, and the relationship between this reference resolution and the model's content capacity $J_M$. (a) Reconstruction PSNR (dB) for datasets with different reference resolutions and using content/style models with different content capacities at $R \in \{2,4\}$ computed on the BrainWeb test volume (300 slices). The black horizontal lines show the baseline reconstruction quality of $L_1$-wavelet compressed sensing (CS-WT). Decreasing the reference resolution consistently drops the reconstruction PSNR, dropping it below the CS-WT baseline at RefRes=1/8$^2$. For a given reference resolution, the reconstruction PSNR is dependent on the model's content capacity $J_M$. (b) Plotting the optimal $J_M$ values for each reference resolution case shows monotonous relationship between them, suggesting that the optimal $J_M$ value can represent a measure of the amount of shared content between the contrasts.}
    \label{fig:sim_jmvsrefres}
\end{figure*}

\subsubsection{Analysis of the content/style prior}
\label{results:sim_cosmoprior_analyses}

\noindent
The content consistency operator at the core of PnP-CoSMo strongly assumes two properties in the content/style model: (a) the content and style representations are sufficiently disentangled, and (b) the content representation estimated from the reference image is sufficiently descriptive. Here, we analyze the influence of the model's content/style disentanglement and its content capacity on the reconstruction quality.

First, we evaluated the effect of perturbing the disentanglement loss weights $\alpha_2$ and $\alpha_3$ (Eq. \eqref{eq:munit_loss}) of MUNIT on the PnP-CoSMo reconstruction quality. Fig. \ref{fig:sim_disent}a shows the reconstruction PSNR for different weight values. 
We first considered different orders of magnitude of values in the range $\alpha_2=\alpha_3 \in [0.001, 10]$ \footnote{Between two given orders of magnitude, we considered three equispaced intermediate values, e.g., between 0.1 and 1, we considered 0.25, 0.5, and 0.75}.
At the lower extreme end with values less than 0.01, poor disentanglement of content and style in the model resulted in poor PnP-CoSMo reconstruction quality. This was expected since PnP-CoSMo relies on the assumption that the content representation is (sufficiently) contrast-independent, which was less enforced at these levels.
At values higher than 1, the training over-emphasized disentanglement while under-emphasizing the GAN and image recovery loss terms, thereby leading to worse preservation of image information in the latent space and thus, to worse reconstructions. At both these extremes, the quality of PnP-CoSMo dropped below a pre-defined baseline reconstruction quality corresponding to $L_1$-wavelet regularized compressed sensing (CS-WT).
We, therefore, consider $\alpha_2=\alpha_3 \in [0.01, 1]$ to be the reasonable range of this setting across datasets.
Analyzing this range further, we explored settings where $\alpha_2 \neq \alpha_3$, as shown in Fig. \ref{fig:sim_disent}b. The optimal combination was found to be $\alpha_2=0.1$ and $\alpha_3=0.25$, producing PSNR of 30.47 dB and 28.12 dB at $R=2$ and $R=4$, respectively. Despite modeling content and style as symmetric and exhaustive categories, we thus find that reconstruction quality can be asymmetrically sensitive to the weighting of their respective disentanglement loss terms. In the subsequent experiments, however, we deviate away from this optimal setting in favor of the theoretical interpretability of the condition $\alpha_2=\alpha_3$ (allowing their interpretation as the level of disentanglement in the model), and set $\alpha_2=\alpha_3=0.1$ (with PSNR of 29.84 dB and 27.33 dB for $R=2$ and $R=4$, respectively) in the following BrainWeb analyses.
Finally, note that the trend in the reconstruction quality over this hyperparameter space remains stable across acceleration values, as observed in the similar pattern between $R=2$ and $R=4$, save for a consistent drop in PSNR in the latter.

PnP-CoSMo relies on a reference image of sufficiently high resolution for guiding the reconstruction. We thus investigated the effect of lowering the reference image resolution on the reconstruction quality and using the content/style model's content capacity (Eq. \eqref{eq:ccapacity}) to explain this effect. We simulated four datasets, denoted as $\text{RefRes}=(1/n)^2$, where $n \in \{1,2,4,8\}$ represents the reference domain downsampling factor. In the $n$=1 case, the T1W/T2W images had the same resolution (as that of the underlying tissue maps), whereas in the subsequent cases, T1W images were blurred to contain only the lower $1/n$ frequency components, while maintaining the same spatial resolution. For each dataset, we trained four content/style models with content capacity $J_M=(1/m)^2$, where $m \in \{1,2,4,8\}$ is the content downsampling factor, which depends on the number of up/downsampling blocks in the networks. For example, the model with $J_M=(1/2)^2$ produced content maps half the spatial resolution of the model with $J_M=1$.
Fig. \ref{fig:sim_jmvsrefres} compares PnP-CoSMo reconstruction quality across these configurations. PFT and CR were disabled here for simplicity. We observe two trends. First, as seen in Fig. \ref{fig:sim_jmvsrefres}a, the reconstruction quality generally decreases with $\text{RefRes}$, eventually dropping below the CS-WT reconstruction baseline. This indicates a decrease in the amount of shared local information contained in the reference contrast. Second, as observed in Fig. \ref{fig:sim_jmvsrefres}b, the \textit{optimal content capacity} of the model decreases with the reference resolution in accordance with the actual amount of the shared information. This implies that optimal content capacity corresponds to the amount of contrast-independent structure the model discovers in the dataset, e.g., the dataset with $\text{RefRes}$=1 had four times the amount of shared structure as the datasets with $\text{RefRes}=1/2^2$, and $\text{RefRes}=1/4^2$, thus explaining the higher effectiveness of the dataset with $\text{RefRes}=1$ in the guided reconstruction. On the other hand, the datasets corresponding to $\text{RefRes}=1/2^2$ and $\text{RefRes}=1/4^2$ had equivalent levels of shared structure (i.e., $J_M=1/2^2=1/4$ times the full-resolution case) despite the lower reference resolution in the dataset with $\text{RefRes}=1/4^2$, suggesting a more complex relationship between the reference resolution and the shared content. In the case with $\text{RefRes}=1/8^2$, the reconstruction quality of PnP-CoSMo reduced to the level of conventional CS-WT reconstruction, suggesting a lower bound on the amount of shared information required for PnP-CoSMo to be effective --- specifically $J_M=1/8^2=1/64$ times the local information of the full-resolution case. 
Thus, the resolution of the reference contrast relative to the target contrast strongly determines the amount of shared content that can be utilized in guided reconstruction. Moreover, prior knowledge of this resolution for a given dataset can be used to estimate the optimal content capacity of the content/style model.

\begin{figure}[tb!]
    \centering
    \includegraphics[width=0.8\textwidth]{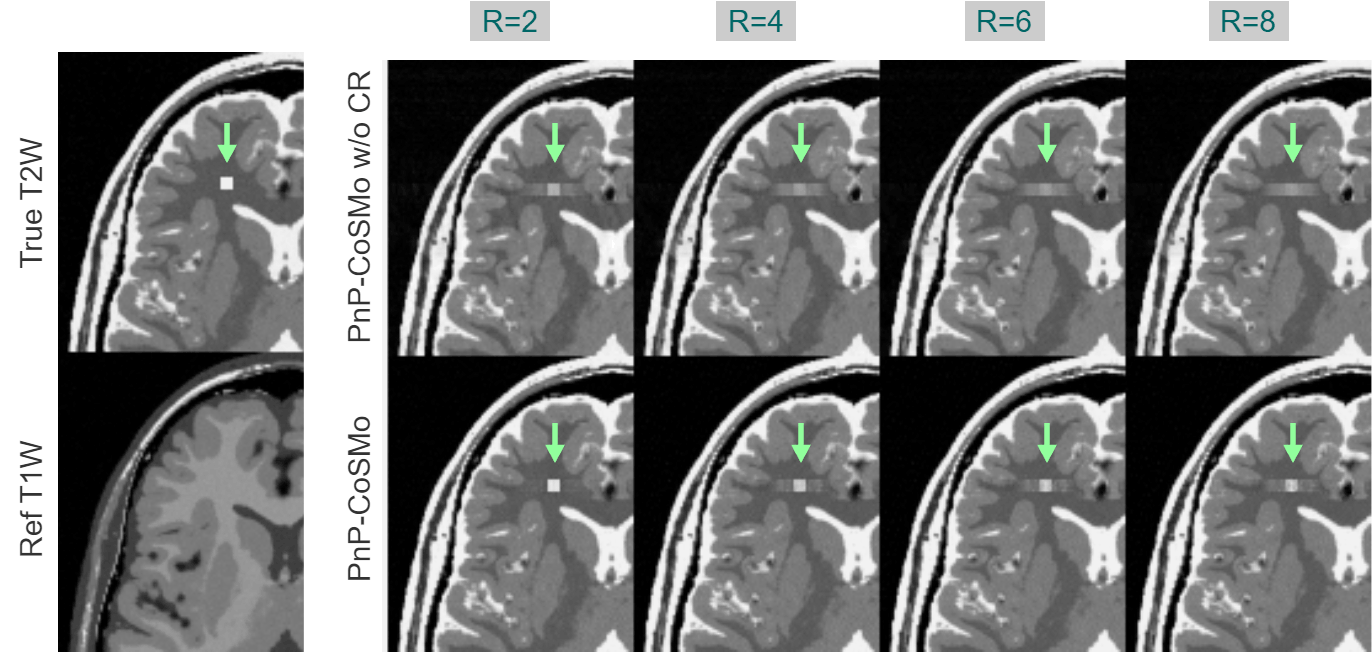}
    \caption{Lesion analysis on synthetic BrainWeb data. Given a contrast-specific structure in the T2W data, such as the simulated lesion here, which is absent in the reference image, our CR module contributes substantially to resolving it.}
    \label{fig:sim_lesion}
\end{figure}
\begin{figure}[tb!]
    \centering
    \includegraphics[width=0.6\textwidth]{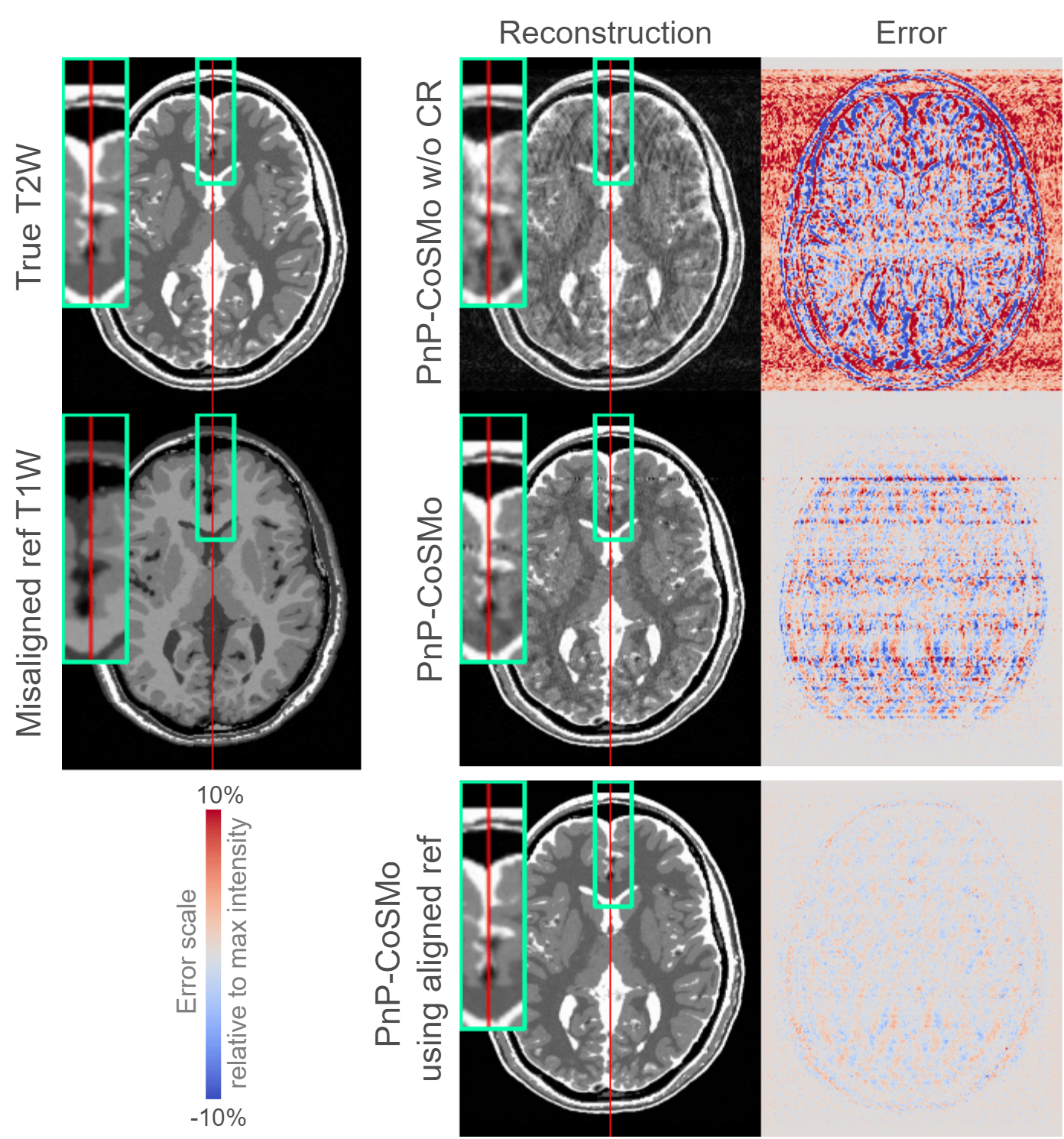}
    \caption{Misalignment analysis on synthetic BrainWeb data. A misaligned reference image (with 2$^{\circ}$ rotation) severely affected the reconstruction when CR was disabled, where enabling CR improved the robustness of the reconstruction process. A PnP-CoSMo reconstruction using an aligned reference is shown for comparison.}
    \label{fig:sim_misalign}
\end{figure}
\begin{figure*}[h!]
    \centering
    \includegraphics[width=0.8\textwidth]{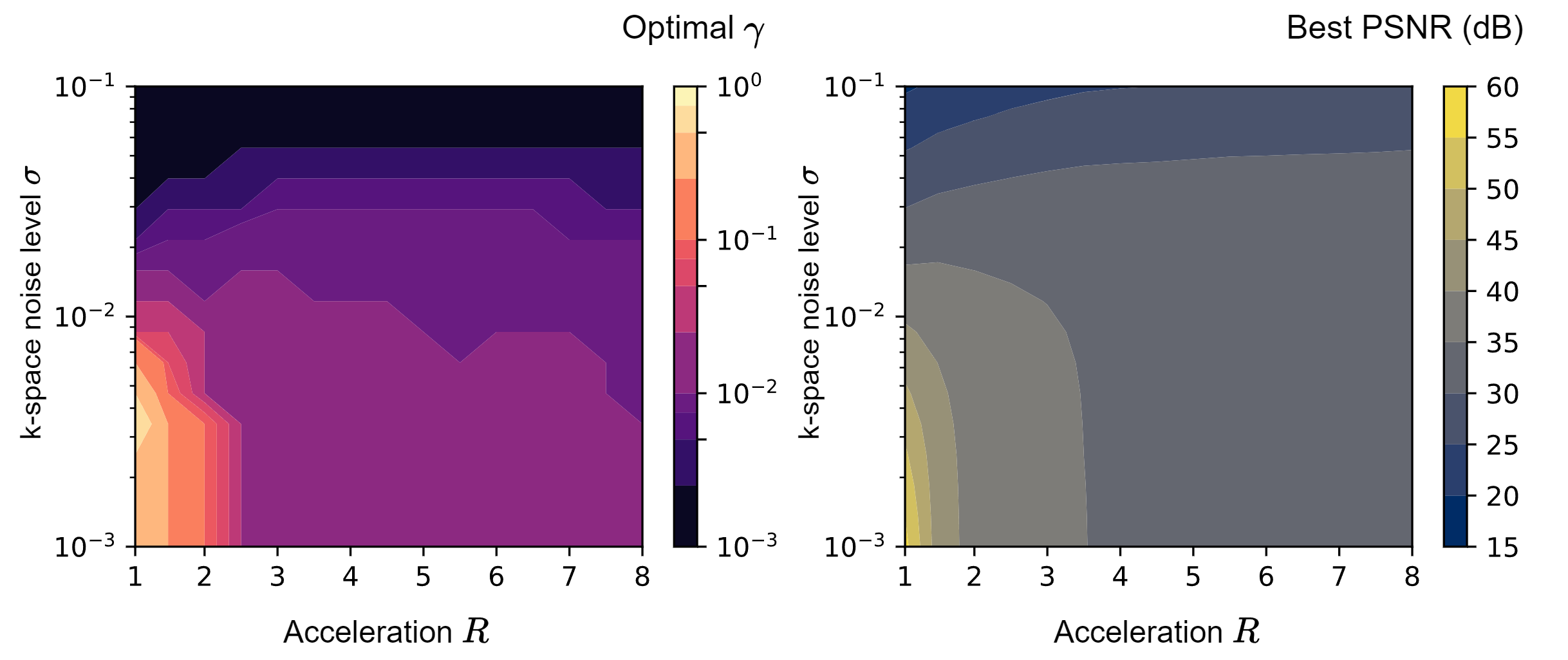}
    \caption{Sensitivity analysis for the CR step size $\gamma$ on synthetic BrainWeb data. The plots show the optimal $\gamma$ (left) and the corresponding best reconstruction PSNR (right) as a function of the k-space acceleration $R$ and noise level $\sigma$. With a decrease in quantity (via undersampling) and in quality (via noise) of the k-space data, we observe that the advantage of the CR module drops as indicated by lower optimal $\gamma$ and lower best PSNR. Moreover, $\gamma$ values in the order of $0.01$ to $0.1$ are optimal over the majority of the k-space configuration space.}
    \label{fig:sim_contentstep}
\end{figure*}
\begin{figure}[tb!]
    \centering
    \includegraphics[width=0.7\textwidth]{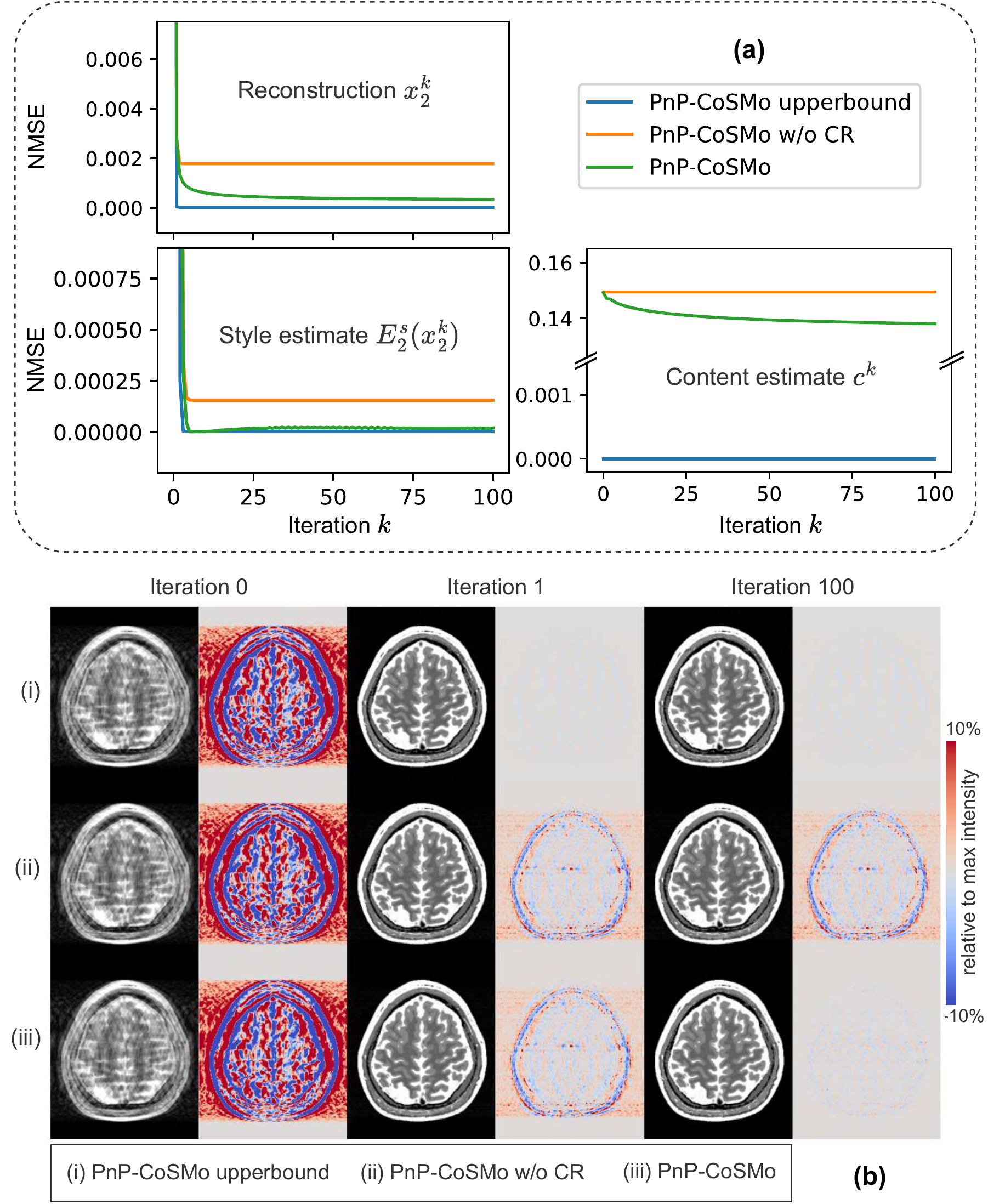}
    \caption{Convergence analysis on synthetic BrainWeb data. (a) Convergence curves for PnP-CoSMo and its two variants at $R$=2. (b) Evolution of the reconstruction shown with the corresponding error maps. In PnP-CoSMo with CR disabled and its upperbound version, i.e., both cases where the content stays constant, the style converges in a single iteration, and so does the reconstruction. However, there is a notable discrepancy in the reference content, which is reflected in the reconstruction error when CR is disabled. Enabling CR leads to an iterative correction of the content discrepancy, leading in turn to the reconstruction approaching the upper-bound quality.}
    \label{fig:sim_convergence}
\end{figure}

\begin{figure*}[tb!]
    \centering
    \includegraphics[width=1\textwidth]{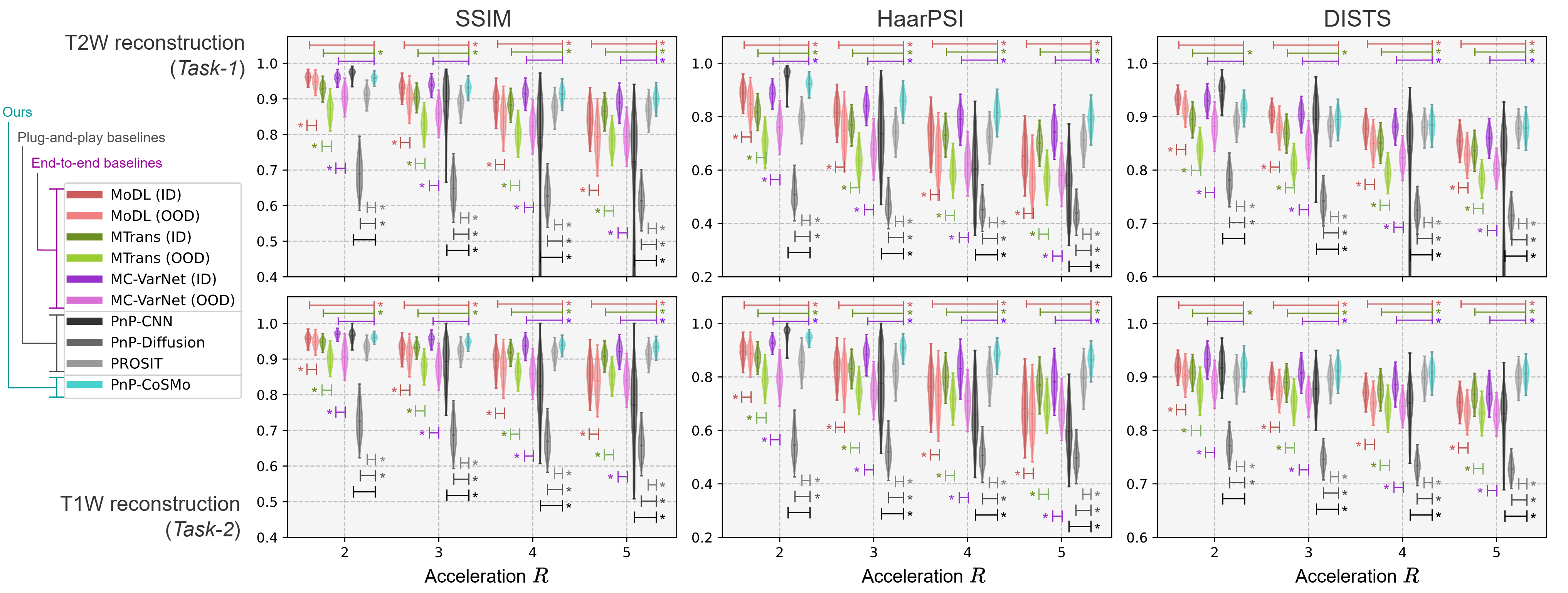}
    \caption{Evaluation plots from the cross-validation on the NYU DICOM dataset for the two reconstruction tasks. ID and OOD refer to in-distribution and out-of-distribution models, respectively. Note that while MTrans, MC-VarNet, PnP-Diffusion, and PROSIT are guided methods, MoDL and PnP-CNN are unguided methods and, hence, did not use the reference image information. Statistically significant comparisons (with $p<0.05$, based on the Wilcoxon signed-rank test) are annotated with $*$. In each task, our PnP-CoSMo showed competitive metrics compared to end-to-end methods and outperformed the other plug-and-play methods. Moreover, unlike the end-to-end methods, which require task-specific k-space training data and which drop in performance on the OOD task, PnP-CoSMo, by design, requires only a single content/style model for both tasks, demonstrating its superior cross-contrast generalizability.}
    \label{fig:nyu_benchmark_metrics}
\end{figure*}

\subsubsection{Analysis of the iterative content refinement procedure}
\label{results:sim_cr_analyses}
\noindent
In the iterative PnP-CoSMo algorithm, the CR update step (Eq. \ref{eq:content_update}) represents a strategy for correcting arbitrary discrepancies between the prior content and the measured k-space. We tested its robustness against two specific discrepancies: (a) contrast-specific structure (e.g., a lesion) present in the target contrast but not in the reference contrast, and (b) spatial misalignment in the reference contrast image. Additionally, we analyzed the influence of the CR step size $\gamma$ on the reconstruction quality and the convergence properties of CR. Here, we used the RefRes-1 dataset and the model with $J_M$=1, which was additionally fine-tuned on two training volumes for 50k iterations. 

For testing the effectiveness of the CR module in resolving structures present exclusively in the target contrast, we simulated a lesion in the T2W image and reconstructed the image with and without the CR module. As shown in Fig. \ref{fig:sim_lesion}, the CR module recovered the lesion fully at $R$=2 and substantially at $R\in\{4,6,8\}$. 

While PnP-CoSMo expects an aligned reference image at reconstruction time, the CR module should, in principle, correct for small misalignments. To test this, we simulated a 2$^{\circ}$ rotation in the reference image and reconstructed the target image at $R$=2 with and without the CR module. Additionally, we consider the case where an aligned reference image is available to PnP-CoSMo as an upper bound. As shown in Fig. \ref{fig:sim_misalign}, although the reconstruction was sensitive to the misalignment in the absence of CR, it significantly improved with CR enabled.

Furthermore, we analyzed the influence of the hyperparameter $\gamma$ in Algorithm \ref{alg:pnp_munit}, which controls the CR step size. 
We simulated k-space accelerations $R \in [1, 8]$ and noise levels $\sigma \in [10^{-3}, 10^{-1}]$, testing $\gamma$ values in the range $ [10^{-3}, 10^{0}]$ at each point in this configuration space.
Fig. \ref{fig:sim_contentstep} shows the optimal $\gamma$ and the corresponding best reconstruction quality as a function of k-space acceleration and noise level. We observe that the optimal $\gamma$ decreased with the amount and quality of k-space data. This was expected as enforcing agreement with the k-space gradually becomes less advantageous, thereby decreasing its contribution to reconstruction quality. Moreover, for the majority of the acceleration-noise space, we observe that the optimal $\gamma$ value lies in the order of $0.01$ to $0.1$, which can hence be considered a reasonable range for setting this hyperparameter.

Finally, we explored the convergence properties of the PnP-CoSMo algorithm, quantifying both the effectiveness and the computational cost of the CR component.
We compared PnP-CoSMo with two other variants of it. The first variant used true content $c^*$ of the ground truth T2W image in the content consistency operator, thereby assuming zero content discrepancy and representing a theoretical upper bound of PnP-CoSMo. The second variant used reference content $\hat{c}$ with the CR module disabled, representing a lower bound of PnP-CoSMo where the non-zero content discrepancy is left uncorrected. Fig. \ref{fig:sim_convergence} shows the convergence curves and intermediate reconstructions. The upper-bound variant converged within a single iteration, the ablated variant converged equally fast, although to a sub-optimal solution. Enabling the CR module closed the gap with the upper bound, although at the cost of slower convergence. It is also observed that the content discrepancy reduced when using CR and that the style estimate always converged within the first few iterations, as we hypothesized in Section \ref{sec:methods-recon}.

\begin{figure*}[tb!]
    \centering
    \includegraphics[width=1\textwidth]{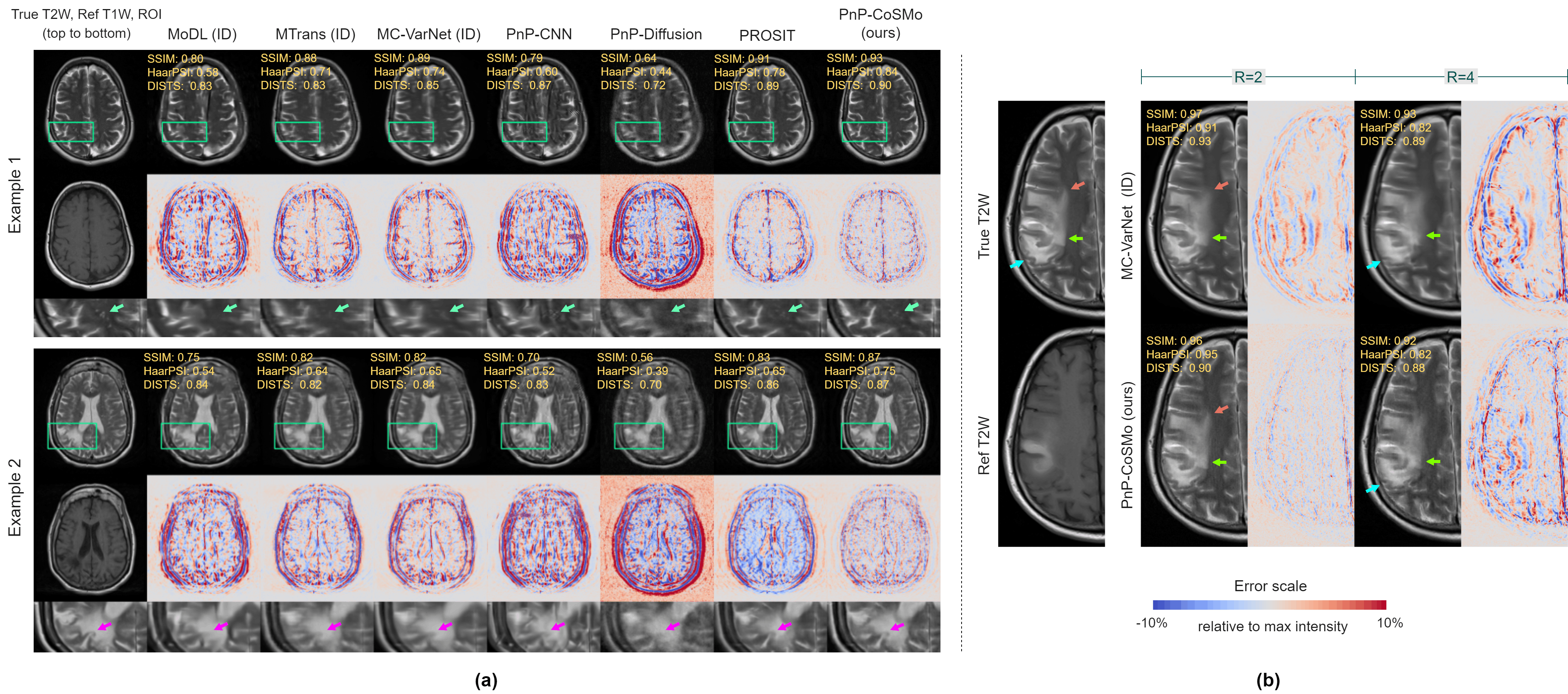}
    \caption{Examples from the NYU benchmark for our Task-1 (i.e., T2W reconstruction). (a) Comparison of PnP-CoSMo with the in-distribution end-to-end models and the plug-and-play baselines at the highest acceleration of $R$=5. PnP-CoSMo produced less blurring and resolved some of the finest details (green and magenta arrows). (b) Comparison between PnP-CoSMo and MC-VarNet at $R$=2 and $R$=4 on a pathological case where many features of the edema are T2W contrast-specific. Again, we observe sharper reconstructions with PnP-CoSMo compared to the end-to-end MC-VarNet, despite often slightly lower metrics.}
    \label{fig:nyu_recons_t2w}
\end{figure*}

\begin{figure*}[tb!]
    \centering
    \includegraphics[width=1\textwidth]{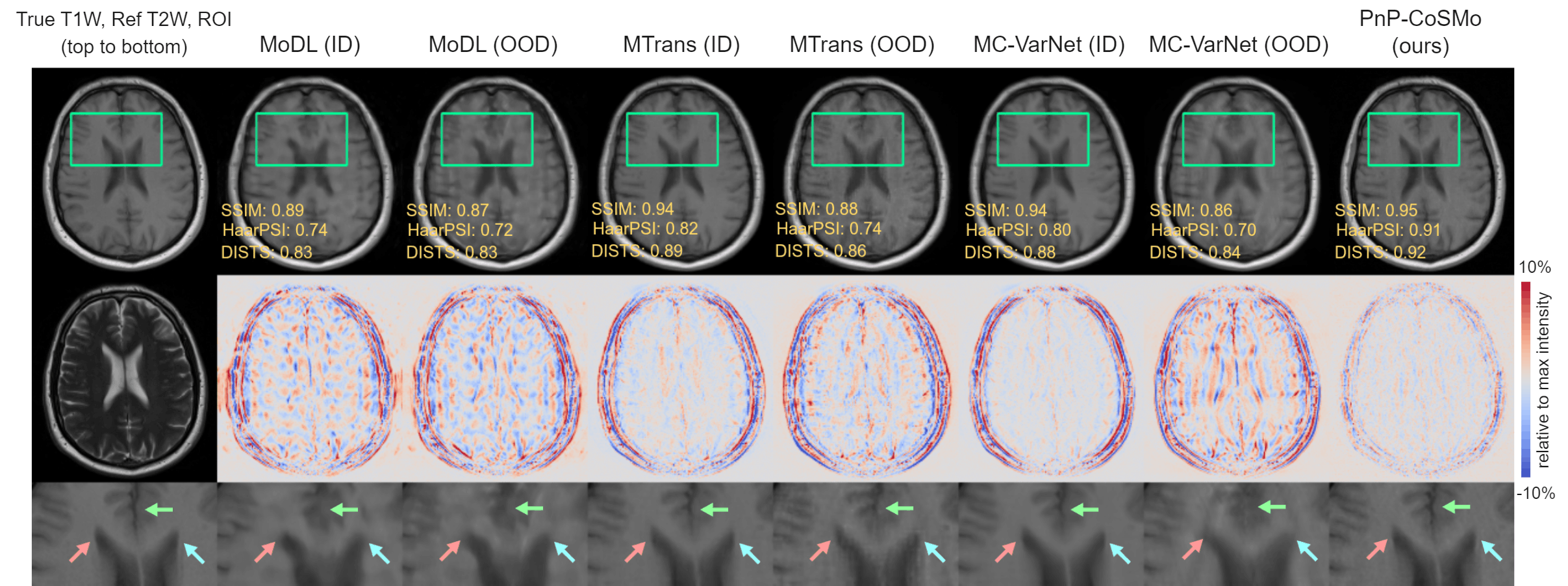}
    \caption{A test example from the NYU benchmark for our Task-2 (i.e., T1W reconstruction) at the highest acceleration $R$=5, demonstrating the cross-contrast generalizability of PnP-CoSMo. The in-distribution (ID) models were trained to solve Task-2 and subsequently applied on Task-2, whereas the out-of-distribution (OOD) ones were trained to solve Task-1 (i.e., T2W reconstruction) and subsequently applied on Task-2. All three end-to-end methods dropped in performance on the OOD case. PnP-CoSMo is immune to such a distribution shift by design. Moreover, it produced sharper reconstructions than the end-to-end models, including the ID case of the latter.}
    \label{fig:nyu_recons_t1w}
\end{figure*}

\subsection{Benchmark on public NYU fastMRI DICOM data}
\label{results:nyu_benchmark}
\noindent
Here we report the findings of the benchmark on the NYU dataset. Let the T2W reconstruction task be denoted as Task-1 and the T1W reconstruction task as Task-2. In Task-1, guided methods use the corresponding T1W image as the reference contrast, and in Task-2, the T2W images are used as the reference. Naturally, unguided methods such as MoDL and PnP-CNN ignore this side information. Fig. \ref{fig:nyu_benchmark_metrics} shows evaluation metrics for PnP-CoSMo and the end-to-end baselines on these two guided reconstruction tasks. 

We observe three main trends. 
First, comparing with the in-distribution reconstruction quality of the end-to-end baselines, PnP-CoSMo outperformed MoDL ($p$$<$0.05 throughout, except for SSIM at $R$=2 and DISTS at $R\in\{2,3\}$ on Task-1, and DISTS at $R$=2 on Task-2) and MTrans ($p$$<$0.05 throughout), and was comparable to MC-VarNet (e.g., with $p$$<$0.05 on Task-2 in terms of DISTS at $R$=3 and SSIM at $R$=4) and even outperformed it ($p$$<$0.05 throughout, except SSIM at $R\in\{2,3,4\}$ and DISTS at $R\in\{2,3\}$ on Task-1, and SSIM at $R\in\{2,3\}$ and DISTS at $R$=2 on Task-2). Compared to the plug-and-play baselines, PnP-CoSMo outperformed PnP-CNN ($p$$<$0.05 throughout, except at $R$=2) and was significantly superior to PnP-Diffusion and PROSIT ($p$$<$0.05 throughout for both).  As shown in Fig. \ref{fig:nyu_recons_t2w}a, PnP-CoSMo could resolve certain fine details even at the highest acceleration of $R$=5, which both end-to-end and plug-and-play baselines struggled with. In the pathological case shown in Fig. \ref{fig:nyu_recons_t2w}b, where many features of the edema were absent in the reference image, PnP-CoSMo produced sharper reconstructions while preserving the structure of the edema better than (the in-distribution) MC-VarNet, despite lower metric values.

Second, considering out-of-distribution generalization, the end-to-end methods consistently dropped in performance on the out-of-distribution tasks ($p$$<$0.05 for all three models on both tasks at all four accelerations). This was expected since end-to-end models are explicitly trained to optimize for a given contrast task and will inevitably fail to generalize across contrasts.

Third, PnP-CoSMo definitively outperformed the out-of-distribution variants of all the end-to-end models ($p$$<$0.05 throughout). Since PnP-CoSMo does not depend on any task-specific k-space training data owing to both its plug-and-play design and the bi-directionality of the content/style model it uses, it is not susceptible to the distribution shift of this kind. This is observed in Fig. \ref{fig:nyu_recons_t1w}.

In summary, PnP-CoSMo produced equivalent or superior reconstruction quality compared to end-to-end methods despite its apparent handicap, i.e., ignoring the k-space training data and not being trained to explicitly solve the reconstruction problem. Moreover, the same feature of PnP-CoSMo makes it immune to contrast-related domain shifts. Finally, PnP-CoSMo was also superior to other plug-and-play methods, both unguided and guided ones, which share a similar design pattern and training data requirements.

\subsection{Benchmark and ablation on in-house LUMC multi-coil data}
\label{results:lumc_benchmark}
\begin{figure*}[tb!]
    \centering
    \includegraphics[width=1\textwidth]{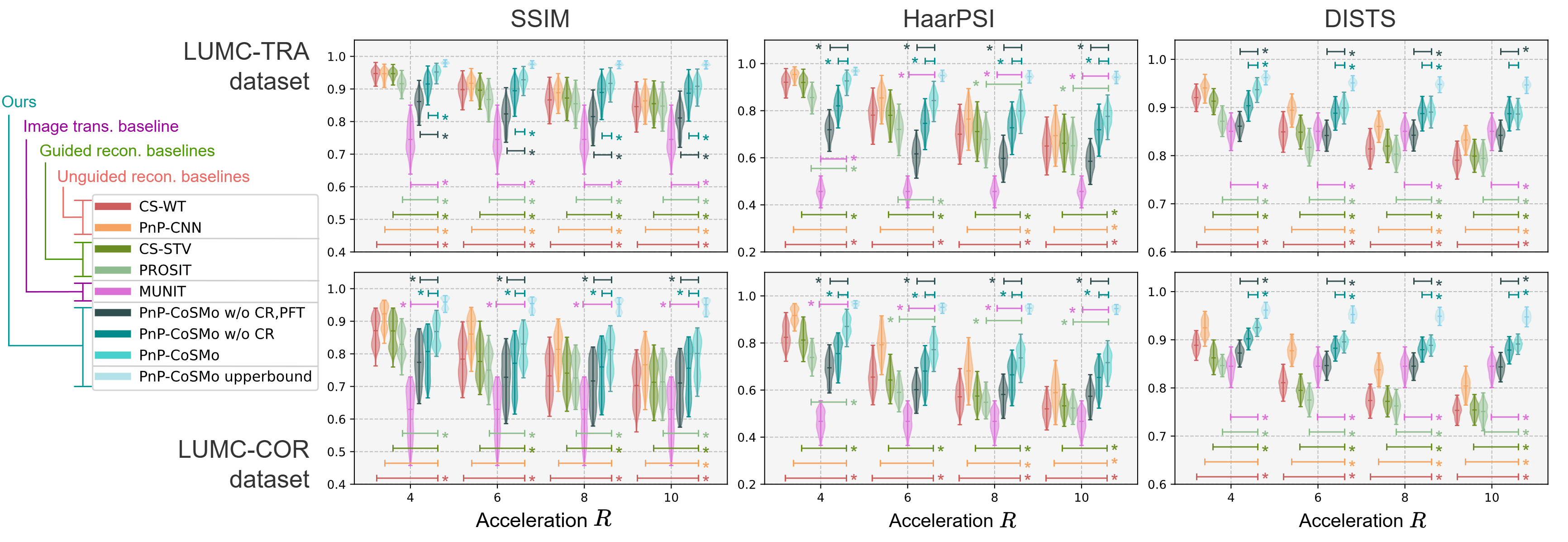}
    \caption{Evaluation plots from the cross-validation on the \textit{recon} subset of LUMC-TRA and LUMC-COR datasets, consisting of 2295 and 412 slices, respectively. Statistically significant comparisons (with $p<0.05$, based on the Wilcoxon signed-rank test) are annotated with $*$. Compared to the unguided reconstruction baselines (CS-WT, PnP-CNN), PnP-CoSMo generally produced higher-quality images, especially at the higher accelerations, showing the advantage of guidance at these accelerations. Compared to the guided reconstruction baselines (CS-STV, PROSIT), reconstructions of PnP-CoSMo were consistently superior, demonstrating its effective use of the reference information. PnP-CoSMo outperformed pure image translation (MUNIT), demonstrating the value of incorporating k-space measurements. Finally, ablating the PFT and CR components from PnP-CoSMo consistently reduced the image quality, thus demonstrating their contribution. The PnP-CoSMo upper bound shows the theoretically achievable quality if the true content could be perfectly estimated.}
    \label{fig:benchmark_metrics_new}
\end{figure*}
\begin{figure*}[tb!]
    \centering
    \includegraphics[width=1\textwidth]{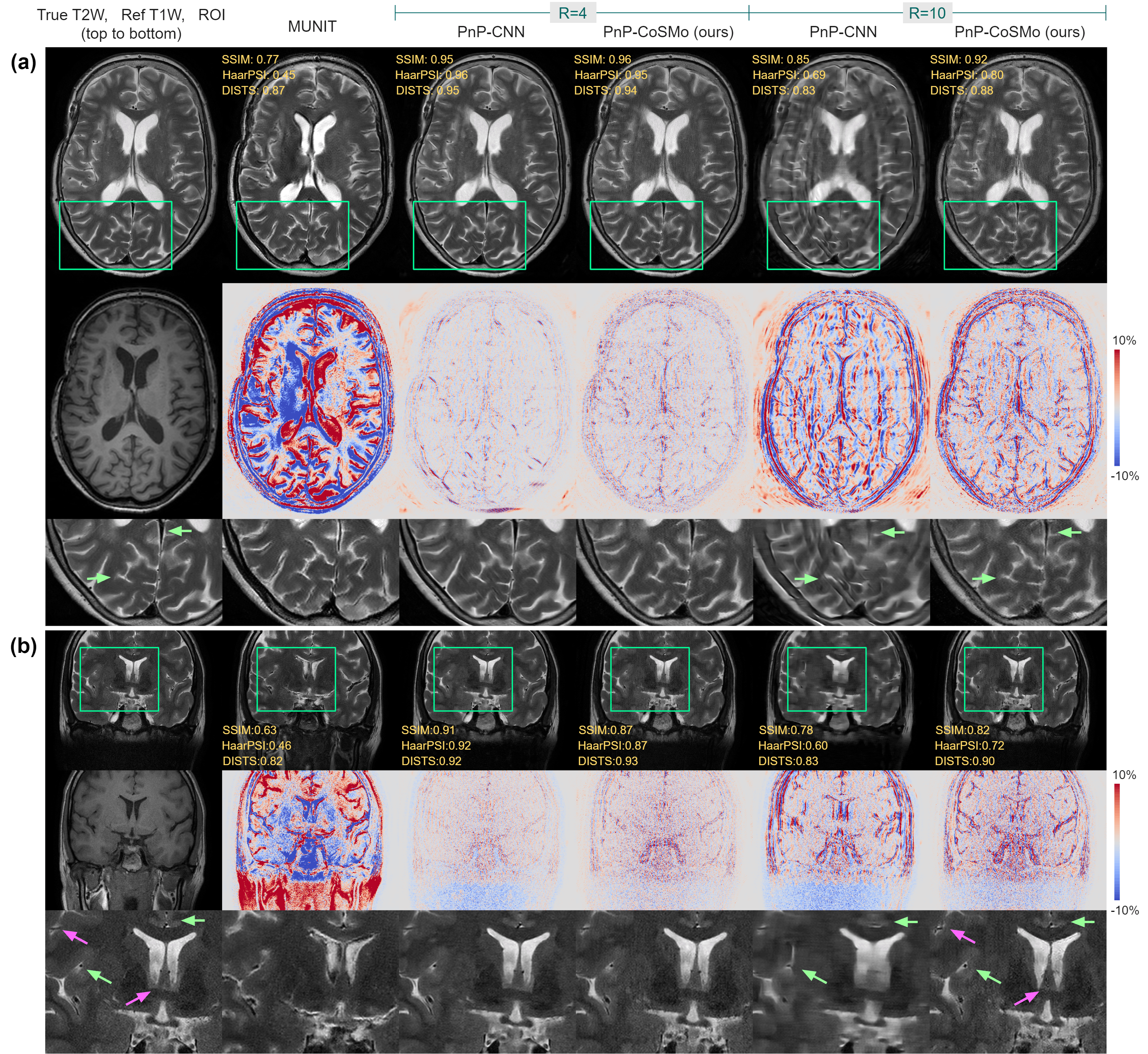}
    \caption{Sample slice from the \textit{recon} set of (a) LUMC-TRA and (b) LUMC-COR, comparing PnP-CoSMo reconstruction with PnP-CNN reconstruction and MUNIT image translation. MUNIT image translation produced severe anatomical defects, which is explained by its lack of access to k-space data. PnP-CoSMo, although comparable to PnP-CNN in quality at $R$=4, produced significantly sharper images at $R$=10, resolving numerous fine details (green arrows). In the LUMC-COR example, a case of model hallucination (magenta arrows) is also observed for PnP-CoSMo at $R$=10, which provides insight into potential failure modes of our algorithm.}
    \label{fig:lumc_recons}
\end{figure*}

\noindent
The LUMC datasets represent a more challenging problem with real-world data constraints. Additionally, in the light of the content capacity analysis (Section \ref{results:sim_cosmoprior_analyses}), we must note that the content capacity $J_M$=1/4 was optimal for the LUMC datasets, compared to NYU fastMRI DICOM data where $J_M$=1 was the optimal content capacity, which reflects a lower effective content information in the LUMC images and hence a greater difficulty of the guided reconstruction problem.

Fig. \ref{fig:benchmark_metrics_new} plots the benchmark metrics for the LUMC test sets, which can be summarized in the following three trends. 
First, pure image translation with MUNIT was worse compared to single-contrast reconstruction, especially at lower acceleration factors (comparing with PnP-CNN, $p$$<$0.05 for all metrics and both datasets). Combining it with L2-regularized reconstruction in PROSIT improved SSIM and HaarPSI ($p$$<$0.05 for both datasets and accelerations), but not necessarily DISTS, suggesting that the available complementary information was not fully utilized. PnP-CoSMo was consistently better than both MUNIT image translation and PROSIT ($p$$<$0.05 throughout for both cases). It also outperformed the conventional guided CS-STV ($p$$<$0.05 throughout, except SSIM at $R$=4 in LUMC-COR). Hence, using both measured k-space and reference scan via PnP-CoSMo was more advantageous than (a) using either one of them and (b) combining both using hand-crafted priors, suggesting that our approach maximally exploits the complementary information. 
Second, both PnP-CNN and PnP-CoSMo produced reconstructions of equivalent quality at lower accelerations, except in terms of SSIM and HaarPSI at $R$=4 and $R$=6 on LUMC-COR, where PnP-CNN was slightly better. At higher acceleration, PnP-CoSMo outperformed PnP-CNN ($p$$<$0.05 for all metrics and both datasets except SSIM at $R$=8 in LUMC-COR). Compared to PnP-CNN, PnP-CoSMo allowed up to 32.6\% more acceleration at given SSIM.\footnote{Based on linear interpolation of LUMC-TRA SSIM, PnP-CoSMo and PnP-CNN allowed $R$=10 and $R$=6.7, which is the maximum difference in $R$, at median SSIM=0.906. Hence, a difference of 32.6\%.} Third, introducing PFT and CR into the ablated PnP-CoSMo improved the reconstructions ($p$$<$0.05 throughout in both cases except in DISTS for the latter at $R$=10 in LUMC-TRA). The contribution of PFT was constant across the acceleration range, whereas that of CR decreased with an increase in acceleration, as expected, since CR depends on the measured k-space data to refine the content.

\begin{figure}[tb!]
    \centering
    \includegraphics[width=0.8\textwidth]{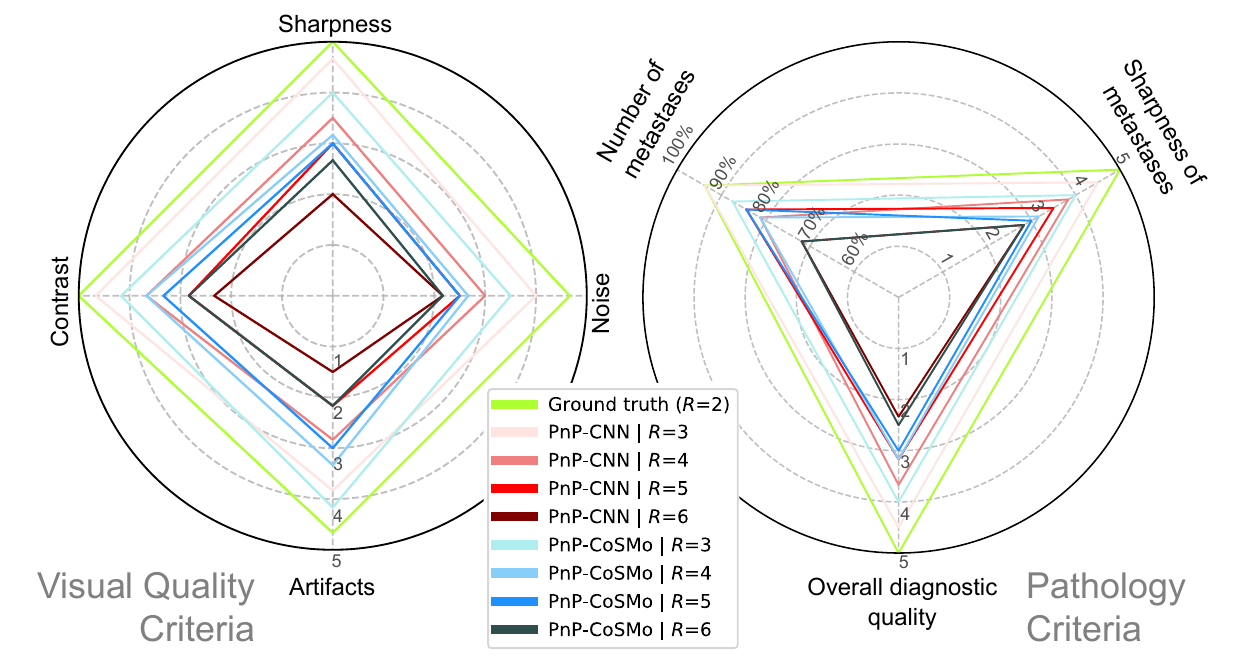}
    \caption{Aggregated scores from our preliminary radiological evaluation. 
    At $R\in\{3, 4, 5\}$, PnP-CoSMo scored at least 3 on the Likert scale (fairly diagnostic quality), except for noise, while PnP-CNN was worse than fairly diagnostic in terms of artifacts at $R\in\{4, 5\}$ and tissue contrast at $R$=5. 
    The main benefit of PnP-CoSMo over PnP-CNN is visible at the highest acceleration of $R$=6, showing a reduction in reconstruction artifacts and an improvement in tissue contrast,  sharpness, and overall diagnostic quality.}
    \label{fig:radeval}
\end{figure}
\begin{figure*}[tb!]
    \centering
    \includegraphics[width=1\textwidth]{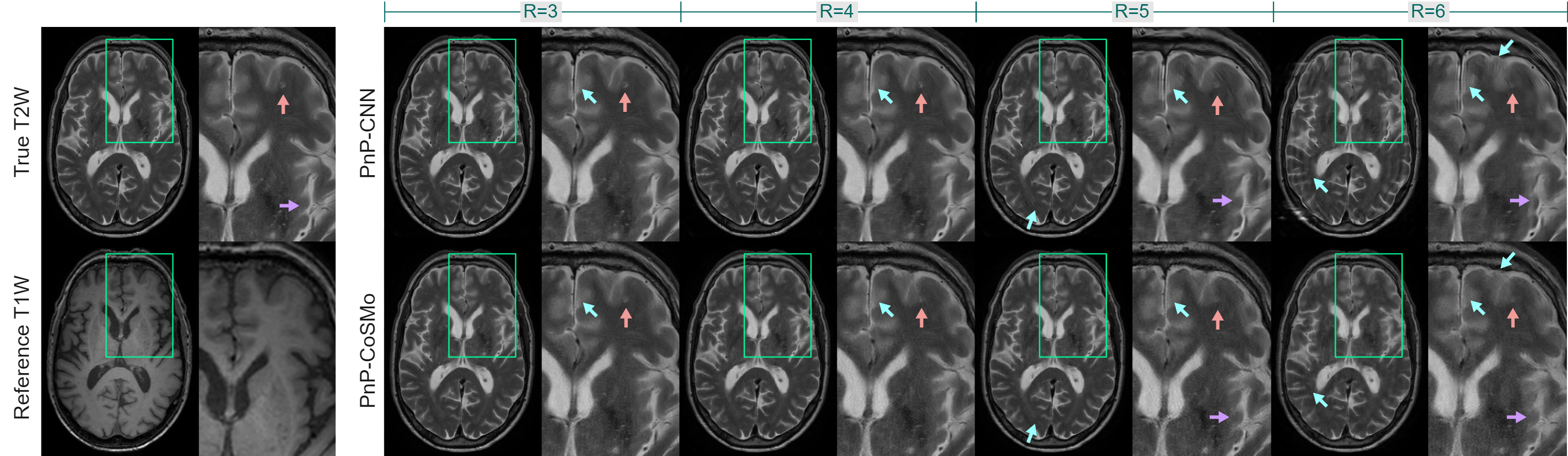}
    \caption{Example slice from an LUMC-TRA subject used in our preliminary radiological evaluation. 
    PnP-CoSMo reconstructions consistently had fewer and less severe undersampling artifacts compared to PnP-CNN (cyan arrows), and less blurring at $R\in\{5, 6\}$ (purple arrows). The pathological hyperintense feature (salmon-pink arrows) is prominently visible at $R$=6 in PnP-CoSMo, unlike in PnP-CNN, where local artifacts and blurring interfere with its visibility.}
    \label{fig:radeval_recons}
\end{figure*}

Fig. \ref{fig:lumc_recons} shows examples from the LUMC-COR and LUMC-TRA datasets representative of the first two trends. We observe that MUNIT image translation produced severe anatomical defects and falsely predicted structure in the lower region of the image, where the ground truth actually contains low signal. At $R$=4, reconstructions of PnP-CoSMo and PnP-CNN were perceptually similar, with the main difference being bluriness in PnP-CNN and mild texture artifacts in PnP-CoSMo. At $R$=10, PnP-CNN reconstruction was non-viable, containing severe blur and artifacts (green arrows). On the other hand, PnP-CoSMo did not severely degrade compared to the $R$=4 case, with fine anatomical structures staying sharply resolved (green arrows). However, a potential failure mode of PnP-CoSMo at high acceleration is observed, namely, subtle localized distortions in the anatomy (magenta arrows).

\subsection{Small-scale qualitative radiological evaluation}
\label{results:radeval}
\noindent
Fig. \ref{fig:radeval} plots the aggregated qualitative scores of our preliminary radiological evaluation, and Fig. \ref{fig:radeval_recons} shows a set of representative example reconstructions. 
At $R\in\{3, 4, 5\}$, PnP-CoSMo produced reconstructions of at least "fairly diagnostic" level (Likert score 3) of image sharpness, artifacts, tissue contrast, and overall diagnostic quality, while enabling the detection of at least 80\% of all metastases. In terms of artifacts, PnP-CoSMo was consistently superior, as seen also in Fig. \ref{fig:radeval_recons} (blue arrows). In terms of tissue contrast, it matched or exceeded PnP-CNN at $R\in\{4, 5, 6\}$ and produced sharper reconstructions at $R\in\{5, 6\}$ (purple arrows in Fig. \ref{fig:radeval_recons}). However, in terms of noise, it was equal or slightly worse. This may be explained by the difference in nature between the denoiser-type model of PnP-CNN and the more complex content/style model of PnP-CoSMo, potentially leading to noise-like grainy texture in the latter, which also interferes with image sharpness and tissue contrast.
In terms of the pathology criteria, while both algorithms scored comparably on the metastasis count, PnP-CoSMo had superior diagnostic quality at $R$=6. The general difference in reconstruction quality at $R$=6 can be seen in Fig. \ref{fig:radeval_recons} where, compared to PnP-CoSMo, the PnP-CNN reconstruction shows severe reconstruction artifacts, blurrier and more distorted features (purple arrows), and a poorly visible hyperintense area of a brain metastasis (salmon-pink arrows). The assessment by the two observers agreed with a Spearman's rank correlation coefficient of 0.81 for the Likert-scale scores and 0.98 for the metastasis count.

In summary, PnP-CoSMo was generally superior to PnP-CNN in terms of reconstruction artifacts and tissue contrast. Even at $R$=6, the highest acceleration considered here, the strong multi-contrast prior in PnP-CoSMo enabled sharper reconstructions with greater tissue contrast, fewer and less severe artifacts, and higher overall diagnostic quality compared to the PnP-CNN baseline. 

\section{Discussion}

\noindent
In this work, we modeled two-contrast MR image data in terms of two latent generative factors ---
\textit{content} representing the contrast-independent structure and \textit{style} representing the contrast-specific variations. We note that our data-driven definitions of content and style, though seemingly related to the MR physics-based representations of quantitative maps and acquisition-related factors, are rather nebulous compared to precisely defined physical concepts. For instance, MR quantitative maps are theoretically objective representations of tissues in terms of physical variables such as relaxation times and are ideally independent of the contrast-generating sequence parameters. On the other hand, our content and style representations are defined statistically in terms of the given two contrasts and are, thus, neither as objective nor as independent in the ideal physical sense. That being said, we would argue that there is potential value in augmenting the data-driven content/style model with MR physical models, e.g., by constraining the learned content to represent physically meaningful anatomical properties, introducing elements of a physical model (e.g., Bloch equations) into the decoder network, etc., to enhance interpretability and reconstruction quality. 

\textit{Optimal content capacity} (Section \ref{sec:methods-model-localitybias}) was defined as a quantity representing the amount of shared local information contained in a two-contrast dataset. 
In terms of this quantity, we analyzed the effect of reference image resolution on reconstruction quality, obtaining a lower bound on the resolution for PnP-CoSMo to be effective (Section \ref{results:sim_cosmoprior_analyses}).
Specifically, this lower limit occurs at a reference resolution $1/8^2$ times lower than the target image resolution, representing a 64-fold reduction in total spatial information.
The optimal content capacity quantity may depend on more fundamental factors, such as MR sequence types, which we did not investigate here. 
While we limited our main experiments to T1W and T2W sequences, PnP-CoSMo is, in principle, applicable to any contrast pair, with image quality limited primarily by the amount of shared structural information. Our supplementary experiments involving the FLAIR and T1WGd contrasts from the NYU dataset, as provided in \ref{appendix:nyu_benchmark_extended}, demonstrate the generality of our approach.
Future work could help empirically determine which sequence pairs are more suitable for guided reconstruction than others, for which large multi-contrast datasets containing matched sets of 3 or more contrasts will be required.

A consequence of our MUNIT-based model design was the model's independence of the unidirectionality inherent in the guided reconstruction problem, which allowed greater generalizability by design (Section \ref{results:nyu_benchmark}).
It is, in principle, also possible to model the content and style for more than two contrasts. 
Given the number of contrasts $N$, content could be defined as the local structure underlying all these contrasts, while style would encode intra-contrast variations of each contrast. This model would then require an (unpaired) image-domain training dataset of the $N$ contrasts. An open practical question is how to effectively scale the model and training complexity to $N$ contrasts.
This would not only expand the generalizability of a given trained content/style model to any combinatorial set of guided reconstruction tasks, but also enable a more accurate estimation of the target contrast's content by utilizing multiple reference contrasts.
To extend the PnP-CoSMo algorithm to utilize multiple reference contrasts, the specific question to investigate is: Given $P$ reference contrasts (where $1<P<N$), what is the optimal aggregation strategy for their contents that maximizes the effectiveness of the content consistency operator?

\textit{Content discrepancy} (Section \ref{sec:methods-recon}) was characterized as a quantity representing the gap between the true content of the target image and our estimation of this content, which is another limiting factor in PnP-CoSMo reconstruction.
In our MUNIT-based network design, the decoders model the (forward) generative process that maps the latent variables (i.e., content and style) to the observable variable (i.e., images). Ideally, the encoders must be a perfect inverse of this forward process. However, estimating the multi-channel contrast-independent content from a single image (using the content encoders) is a challenging (and perhaps ill-conditioned) inverse problem. Hence, irreducible residual errors are to be expected in the content estimated from the reference image.
Moreover, the reference image may contain unfavorable deviations from the target image, such as spatial misalignment or artifacts, which exacerbate the problem.
Therefore, content discrepancy is expected to be a non-zero quantity. We proposed two complementary strategies to minimize its effect on the reconstruction. 
While paired fine-tuning improves the model using limited amounts of paired image-domain training data, content refinement (CR) aims to correct the remaining discrepancy at reconstruction time using the measured k-space. A drawback of the CR module, however, is slower convergence (Section \ref{results:sim_cr_analyses}). 
For example, the run-time for PnP-CoSMo on the GPU was approximately 17 seconds for a slice of matrix size 349$\times$284, compared to 6 seconds for CS-WT, 1.4 seconds for PnP-CNN, and less than 0.1 seconds for MC-VarNet.
This latency issue could be mitigated by approximating the iterative process with an unrolled network design, although potentially at a cost of the generalizability offered by the plug-and-play design. The resulting trade-off between model latency and generalizability would be a further avenue for future exploration.

In terms of training data requirements, PnP-CoSMo can utilize larger amounts of available image-domain data and is applicable in situations where end-to-end methods are infeasible. In cases where end-to-end methods are indeed feasible, PnP-CoSMo is a competitive alternative with superior generalizability. In our NYU fastMRI benchmark (Section \ref{results:nyu_benchmark}), PnP-CoSMo was equivalent or superior to state-of-the-art unguided (MoDL) and guided (MC-VarNet and MTrans) networks in terms of numerical metrics and perceptual quality, especially at higher accelerations, e.g., at $R$=5 with a difference of 0.01 SSIM, 0.04 HaarPSI, and 0.02 DISTS on average over MC-VarNet in Task-1 and of 0.01 SSIM, 0.08 HaarPSI, and 0.04 DISTS in Task-2, despite PnP-CoSMo \textit{completely ignoring the k-space training data and not being explicitly optimized for any reconstruction task in the training phase}. Additionally, PnP-CoSMo is, by design, immune to cross-contrast distribution shifts as opposed to end-to-end methods. This was demonstrated, e.g., in comparison with MC-VarNet, which suffered a difference of up to 0.10 SSIM, 0.17 HaarPSI, and 0.06 DISTS on average when applied on out-of-distribution tasks. Compared to other plug-and-play methods, which share a similar decoupled design pattern and the image-only training data requirement as PnP-CoSMo, our method produced sharper and higher-fidelity reconstructions, e.g., with a maximum difference of 0.05 SSIM, 0.13 HaarPSI, and 0.03 DISTS on average over PROSIT in Task-1 and of 0.03 SSIM, 0.10 HaarPSI, and 0.02 DISTS in Task-2.
Moreover, the \textit{unpaired} image-domain pre-training of the content/style model expands the practical applicability of PnP-CoSMo, e.g., in the case of LUMC-TRA, where the T1W/T2W image data imbalance was considerable (Table \ref{tab:lumc_splits}). 
In our LUMC multi-coil benchmark, where no raw k-space data was available for training, we limited our baselines to feasible algorithms given the data constraint. As future work, a more comprehensive study should be conducted with a broader range of methods and evaluation criteria, e.g., comparing generalizability across accelerations and sampling patterns with multi-contrast unrolled networks, given a fixed budget of paired training data. Contrary to the training stage, we strictly assumed spatially aligned reference images available in the reconstruction stage.
The CR module, which can implicitly correct minor registration errors (as shown in Section \ref{results:sim_cr_analyses}), would likely break down at the empirically observed levels of patient motion. A potential solution is to incorporate an online registration step to explicitly and efficiently correct for arbitrary inter-scan motion, thereby further improving practical applicability.

In terms of reconstruction quality on multi-coil data, we observed in Section \ref{results:lumc_benchmark} that the true advantage of PnP-CoSMo was observed at the high acceleration factors ($R\in\{8, 10\}$) where the k-space data is scarce, and the reference information becomes more valuable. For example, at $R$=10 and compared to PnP-CNN, PnP-CoSMo showed an improvement of 0.04 SSIM, 0.08 HaarPSI, and 0.05 DISTS on LUMC-TRA and of 0.02 SSIM, 0.12 HaarPSI, and 0.09 DISTS on LUMC-COR.
However, the risk of model hallucinations increases at these extreme accelerations, raising concerns about the accuracy of the anatomy represented in the image. 
We hypothesize that, as a consequence of content/style decomposition, it may be possible to leverage the contrast-independence property of the content to: (a) automatically detect local hallucinations in the reconstruction, and (b) define a corrective process in combination with an adaptive sampling strategy to minimize them. This is yet another topic for future work.
At lower accelerations, on the other hand, we observed that PnP-CoSMo often produced marginally lower metrics compared to PnP-CNN, e.g., in LUMC-COR. This was also observed in our qualitative radiological evaluation on LUMC-TRA samples (Section \ref{results:radeval}), with PnP-CoSMo reconstructions at $R\in\{3, 4\}$ containing noise-like grainy texture, which also interfered with image quality criteria such as sharpness. At first glance, this is counterintuitive since PnP-CoSMo has access to additional side information and should, in principle, perform at least as well as PnP-CNN. 
However, this can be explained by the fact that at lower accelerations, the quality of the content/style model $M$ becomes the primary limiting factor of reconstruction quality. This is because, unlike the CNN denoiser (Eq. \eqref{eq:pnpcnn}) of PnP-CNN, the content consistency operator $g_M(\cdot, c)$ (Eq. \eqref{eq:munit_op2}) of PnP-CoSMo radically alters the image, making this operation more sensitive to the model quality. The model quality is, in turn, influenced by the training dataset and the model design. First, in terms of training data, the T1W scans in LUMC datasets had an in-plane resolution 1.3-2.5 times lower than that of the T2W scans, limiting the quality of the model (as indicated by its low optimal content capacity) and the overall value of guidance. In the long term, this may be solved by using 3D reference and target sequences of matched resolution to optimize guidance. Second, in terms of model design, correcting for content discrepancy using the CR module relies on computing gradients through the decoder $G_2$ (Eq. \eqref{eq:content_update}). If $G_2$ is highly non-linear and the reference content is not sufficiently close to the optimum, CR may converge to a sub-optimal point leading to grainy texture in the reconstruction, e.g., as seen in Fig. \ref{fig:lumc_recons} at $R$=4. In our proof-of-concept, we used the identical model architecture as proposed in \citet{huang2018multimodal}, with minor modifications. Improved designs that are more conducive to the gradient-based CR procedure may exist. Hence, future efforts will focus considerably on improving the content/style model using specialized architecture and regularization.

Finally, our small-scale qualitative radiological study demonstrated that PnP-CoSMo produced reconstructions with greater sharpness and tissue contrast, fewer artifacts, and higher overall diagnostic quality compared to the single-contrast PnP-CNN at $R$=6 --- a high acceleration factor based on clinical standards.
As a supplementary component to our LUMC multi-coil benchmark, this small-scale qualitative evaluation merely provides a preliminary indication of the potential benefit of our approach in the context of a specific clinical problem. Given the limited scale of our current evaluation, however, a large-scale clinical validation involving a larger sample size and multiple radiologists must be conducted to robustly measure the clinical utility of our approach in quantitative terms. Such clinical validation was beyond the scope of our current technical work and is, hence, a topic for a clinically-focused future work.

\section{Conclusion}
\noindent
In this work, we introduced PnP-CoSMo, a modular approach to multi-contrast reconstruction that combines content/style modeling with iterative reconstruction, offering the reconstruction quality of state-of-the-art end-to-end methods with improved cross-contrast generalizability despite not requiring any k-space training data.
At its core is the content consistency operation, which provides regularization at the semantic level of the image content. We defined two quantities that determine the efficacy of this operation, namely optimal content capacity and content discrepancy, and provided several ways of maximizing this efficacy.
On real-world clinical data, PnP-CoSMo provided up to 32.6\% more acceleration over PnP-CNN for given SSIM, enabling sharper reconstructions at high accelerations. To progress from our current proof-of-concept towards a practical implementation, future work will focus mainly on improving and extending the content/style model, reducing reconstruction latency, addressing the problem of model hallucinations, incorporating online registration, and conducting a large-scale radiological study.

\section*{CRediT authorship contribution statement}
\noindent
\textbf{Chinmay Rao:} Conceptualization, Methodology, Data curation, Software, Investigation, Validation, Formal analysis, Visualization, Writing – original draft.
\textbf{Matthias van Osch:} Funding acquisition, Supervision, Conceptualization, Validation, Writing – review \& editing.
\textbf{Nicola Pezzotti:} Funding acquisition, Supervision, Conceptualization, Methodology, Validation.
\textbf{Jeroen de Bresser:} Investigation, Validation.
\textbf{Mark van Buchem:} Investigation, Validation.
\textbf{Laurens Beljaards:} Data curation, Software.
\textbf{Jakob Meineke:} Resources, Software.
\textbf{Elwin de Weerdt:} Conceptualization, Resources.
\textbf{Huangling Lu:} Investigation.
\textbf{Mariya Doneva:} Supervision, Conceptualization, Methodology, Validation, Writing – review \& editing.
\textbf{Marius Staring:} Funding acquisition, Project administration, Supervision, Conceptualization, Methodology, Validation, Writing – review \& editing.

\section*{Declaration of competing interest}
\noindent
The authors declare the following financial interests or personal relationships which may be considered as potential competing interests:
Chinmay Rao and Laurens Beljaards declare the grant and research support they receive from Philips.
Nicola Pezzotti, Jakob Meineke, Elwin de Weerdt, and Mariya Doneva declare their employment at Philips.
If there are other authors, they declare that they have no known competing financial interests or personal relationships that could have appeared to influence the work reported in this paper.

\section*{Acknowledgments}
\noindent
This work is part of the project ROBUST: Trustworthy AI-based Systems for Sustainable Growth with project number KICH3.LTP.20.006, which is (partly) financed by the Dutch Research Council (NWO), Philips Research, and the Dutch Ministry of Economic Affairs and Climate Policy (EZK) under the program LTP KIC 2020-2023.

\section*{Data availability}
\noindent
The source code for PnP-CoSMo is publicly available on GitHub at \href{https://github.com/cnmy-ro/pnp-cosmo}{github.com/cnmy-ro/pnp-cosmo}. The authors do not have permission to share the in-house MRI data.

\appendix
\setcounter{figure}{0}
\section{NYU Benchmark on Additional Contrasts}
\label{appendix:nyu_benchmark_extended}

\noindent
While the main body of experiments focused on T1W and T2W contrasts, we extend the NYU benchmark in this section by evaluating PnP-CoSMo on two additional contrasts, namely T1WGd and FLAIR. 
To this end, we constructed two separate datasets: (a) a set of 447 subjects with paired T1WGd and T2W, and (b) a set of 392 subjects with paired T2W and FLAIR. The T1WGd scans were 2D axial gradient echo acquisitions with a median in-plane FOV of 220$\times$220 mm$^2$, in-plane voxel size of 0.69$\times$0.69 mm$^2$, slice thickness of 5 mm, FA=70$^{\circ}$, TR=400 ms, and TE=4.24 ms. The FLAIR scans were 2D spin echo axial acquisitions with a median in-plane FOV of $220\times$220 mm$^2$, in-plane voxel size of 0.43$\times$0.43 mm$^2$, slice thickness of 5 mm, FA=150$^{\circ}$, TR=900 ms, TE=94 ms, ETL=16, and TI=2500 ms. The corresponding two contrasts in each dataset were then registered.
We used the two datasets to evaluate our method on two separate guided reconstruction tasks: (a) T1WGd-guided T2W reconstruction and (b) T2W-guided FLAIR reconstruction. As baselines, we used the strongest of our end-to-end and plug-and-play baselines, namely MC-VarNet and PROSIT, respectively. Using an experimental setup identical to that used in Section \ref{expt:nyu}, we conducted a three-fold cross-validation on both datasets using the same training and testing setups, except that no out-of-distribution cross-contrast analysis was performed here. All models were newly initialized and trained separately for these two reconstruction tasks.

\begin{figure*}[tb!]
    \centering
    \includegraphics[width=1\textwidth]{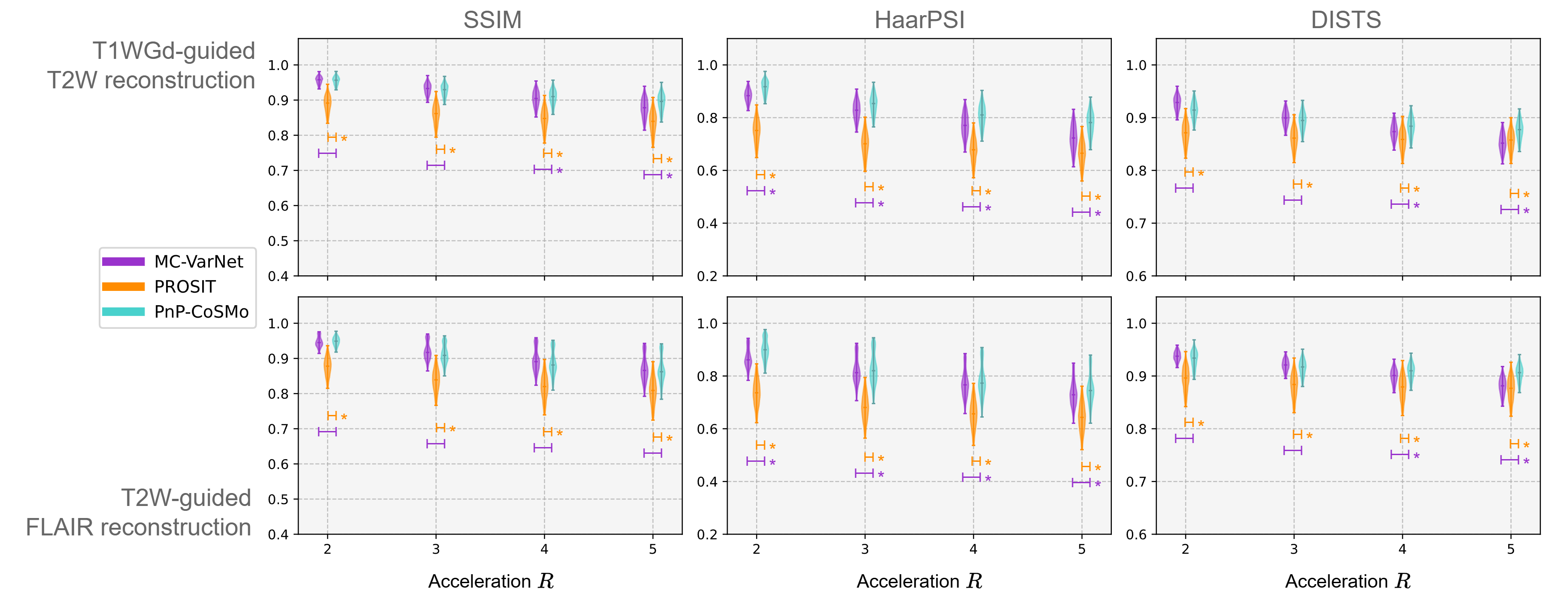}
    \caption{Evaluation plots from the cross-validation on the NYU data for two new reconstruction datasets. Statistically significant comparisons (with $p<0.05$, based on the Wilcoxon signed-rank test) are annotated with $*$. PnP-CoSMo consistently outperformed PROSIT on both datasets. It was comparable to MC-VarNet while outperforming the latter in several cases, despite not requiring k-space training data.}
    \label{fig:nyu_benchmark_metrics_expanded_mediarev3}
\end{figure*}

\begin{figure*}[tb!]
    \centering
    \includegraphics[width=1\textwidth]{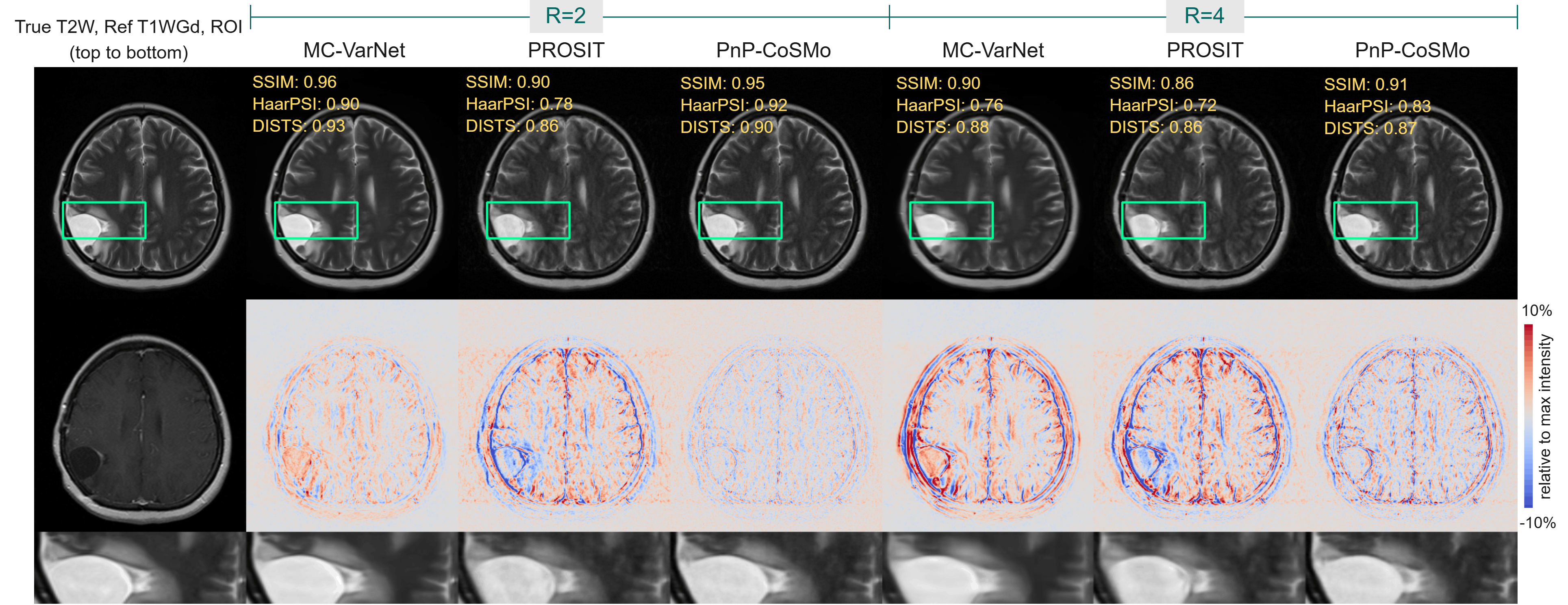}
    \caption{An example set of T2W reconstructions from the NYU dataset with T1WGd guidance at $R \in \{2,4\}$. PnP-CoSMo reconstructions are sharper than MC-VarNet's, with noticeably reduced blurring at $R=4$. Compared to PROSIT, PnP-CoSMo produced less artifacts and distortions, as observed at both $R=2$ and $R=4$.}
    \label{fig:nyu_recons_t1gdt2}
\end{figure*}
\begin{figure*}[tb!]
    \centering
    \includegraphics[width=1\textwidth]{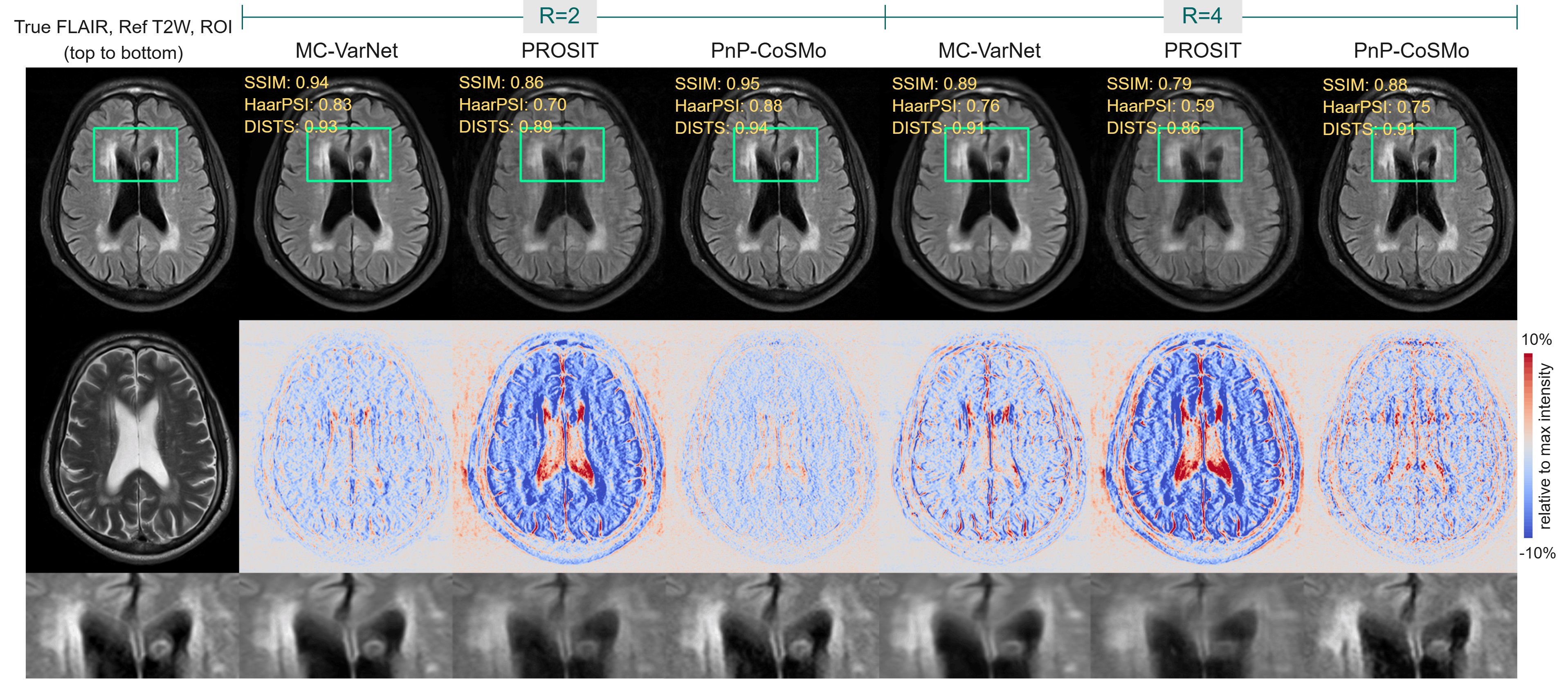}
    \caption{An example set of FLAIR reconstructions from the NYU dataset with T2W guidance at $R \in \{2,4\}$. Reconstructions from PnP-CoSMo are sharper than those from both MC-VarNet and PROSIT at both accelerations. At $R=2$, the pathological structure is more prominently resolved by PnP-CoSMo.}
    \label{fig:nyu_recons_flairt2}
\end{figure*}

Fig. \ref{fig:nyu_benchmark_metrics_expanded_mediarev3} plots the metrics for the two datasets. PnP-CoSMo consistently outperformed PROSIT on both tasks across all four accelerations in terms of all three metrics ($p<0.05$ throughout). Against the end-to-end MC-VarNet, Pnp-CoSMo was comparable, while outperforming it in the majority of the cases ($p<0.05$ at all accelerations in terms of HaarPSI and at $R \in \{4,5\}$ in terms of DISTS on both datasets, and at $R \in \{4,5\}$ in terms of SSIM on T1WGd-guided T2W reconstruction). As seen in Figs. \ref{fig:nyu_recons_t1gdt2} and \ref{fig:nyu_recons_flairt2}, PnP-CoSMo produced reconstructions that were sharper than those from MC-VarNet and contained fewer distortions than PROSIT. 

We have hereby demonstrated the applicability of the PnP-CoSMo algorithm across different contrasts. PnP-CoSMo maintained its superior performance compared to a strong plug-and-play baseline (up to 0.07 higher SSIM, 0.17 higher HaarPSI, and 0.04 higher DISTS than PROSIT). It also maintained a competitive performance compared to a strong end-to-end reconstruction model (up to 0.01 higher SSIM, 0.05 higher HaarPSI, and 0.02 higher DISTS than MC-VarNet) despite requiring no k-space training data.

\bibliographystyle{elsarticle-harv}
\bibliography{references.bib}
\end{document}